\documentclass[final]{svjour3}

\usepackage{graphicx}
\usepackage{bmpsize}
\usepackage{amsmath}
\usepackage{amsfonts}
\usepackage{amssymb}
\usepackage{latexsym}
\usepackage{graphicx}
\usepackage{color}
\usepackage{mathrsfs}
\usepackage{slashed}
\usepackage{soul}
\usepackage{array}
\usepackage{cite}
\usepackage{placeins}
\usepackage{hyperref}

\usepackage{enumerate}
\usepackage{multirow}
\usepackage[table]{xcolor}
\usepackage{colortbl}

\usepackage{float}

\usepackage{axodraw4j}
\usepackage[text={7in,10in}]{geometry}

\usepackage{comment}
\includecomment{toexclude} 
\excludecomment{addsections} 

\numberwithin{equation}{section}

 \def\be   {\begin{equation}}   \def\ee   {\end{equation}}
       \def\ea   {\end{array}}
 \def\bea  {\begin{eqnarray}}   \def\eea  {\end{eqnarray}}
 \def\bean {\begin{eqnarray*}}  \def\eean {\end{eqnarray*}}

\begin{document}
\hfill\textit{DO-TH 17/14}


\begin{center}
{\Huge
Leptonic Flavor Structure in the Brane Shifted Extra Dimensional Seesaw Mechanism
}
\\ [2.5cm]
{\large{\textsc{ 
Mathias Becker\footnote{\textsl{mathias.becker@tu-dortmund.de}} ,
Heinrich P\"as\footnote{\textsl{heinrich.paes@tu-dortmund.de}} 
}}}
\\[1cm]

\large{\textit{
Fakult\"at f\"ur Physik, Technische Universit\"at Dortmund,\\
44221 Dortmund, Germany
}}
\\ [2 cm]
{ \large{\textrm{
Abstract
}}}
\\ [1.5cm]
\end{center}

We discuss the leptonic flavor structure generated by a brane shifted extra dimensional seesaw model with a single right handed neutrino in the bulk. \\
In contrast to previous works, no unitarity approximation for the $3 \times 3$ submatrix has been employed. This allows to study phenomenological signatures such as lepton flavor violating decays. \\
A strong prediction of the model, assuming CP conservation, are the ratios of flavor violating charged lepton decay and Z decay branching ratios which are correlated with the neutrino mixing angles and the neutrino mass hierarchy. Furthermore, it is possible to obtain branching ratios for $\mu \rightarrow e \gamma$ close to the experimental bounds even with Yukawa couplings of order one.

\def\thefootnote{\arabic{footnote}}
\setcounter{footnote}{0}
\pagestyle{empty}

\newpage
\pagestyle{plain}
\setcounter{page}{1}

\section{Introduction}\label{sec:Introduction}
In the last decade compactified large extra dimensions (LED) attracted a lot of attention \cite{ArkaniHamed:1998rs,Randall:1999vf}, by providing an attractive possibility to solve the hierarchy problem. This can be achieved by allowing Standard Model (SM) singlets, e.g. gravitons, to propagate in spatial extra dimensions leading to a suppression of the Planck scale by a volume factor of the extra dimensions. \\
Since a right handed neutrino is a SM singlet it could also be allowed to propagate in the extra dimension, resulting in a suppression of the Yukawa coupling to the left handed neutrino, and thereby, suppressing the neutrino mass \cite{ArkaniHamed:1998vp}. Additionally, if a right handed neutrino feels the extra dimensions, an infinite tower of Kaluza-Klein excitations with masses $\sim R^{-1}$ appears when integrating out the extra dimensions, resulting in an additional suppression of the neutrino mass by an extra dimensional variant of the type I seesaw mechanism, which was investigated e.g. in \cite{Dienes:1998sb,Bhattacharyya:2002vf,Grossman:1999ra}. \\
In this paper, we explain the observed active neutrino mixing within a minimal extra dimensional extension of the SM, where only one right handed neutrino field is introduced, which can propagate in one extra dimension while gravity is allowed to feel a larger number of extra dimensions. Furthermore, the brane where the SM particles and interactions are located is shifted away from the fix points of the $S_1 /  \mathcal{Z}_2$ orbifold.     
Without this brane shift the model is only capable of generating one neutrino mass difference. Consequently, the brane shift is necessary to generate a realistic result.  A similar setup was discussed with only one generation of neutrinos and a focus on neutrinoless double beta decay \cite{Bhattacharyya:2002vf} or on leptogenesis \cite{Pilaftsis:1999jk} while in \cite{Ioannisian:1999cw} lower limits on the fundamental scale of gravity were derived. A systematic study of right handed neutrinos with a bulk mass term propagating within one flat extra dimension is given in \cite{Lukas:2000rg}.  \\
An important consequence of the active neutrino mixing with sterile neutrinos is that the resulting effective three by three mixing matrix of the active neutrinos is not unitary anymore. This leads to some phenomenological consequences, e.g. in rare lepton decays \cite{Ioannisian:1999cw,Antusch:2003kp, Antusch:2014woa}. \\
The paper is structured as follows: In section \ref{sec:setup} the general setup is introduced and the complete mass matrix for the active neutrinos and the Kaluza Klein excitations is derived. In section \ref{sec:mass+pmns} this mass matrix is analyzed. By employing some approximations, the mixing matrix for the neutrinos in a non-unitarity violating limit is derived and used to constrain the parameter space of the model. Finally, in section \ref{sec:UVandPheno}, the unitarity violation of the system is investigated in more detail and the resulting effects on lepton flavor violating decays are studied. 
\section{Setup}\label{sec:setup}
\begin{figure}
\centering
\includegraphics[width=8cm]{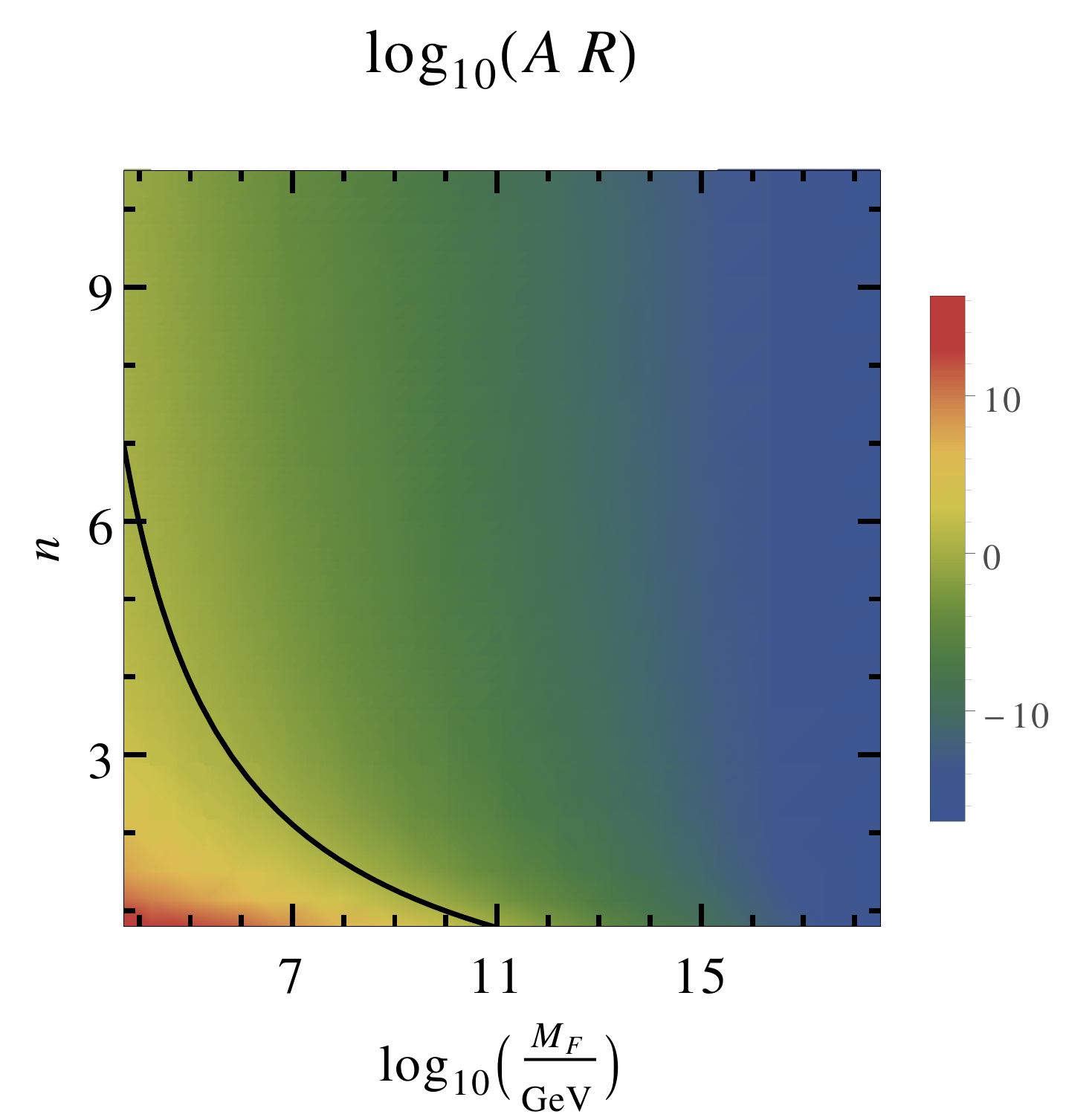}
\caption{$\log_{10} \left( A^l R \right)$ in dependence of the number of extra dimensions $n$ and the new fundamental scale of gravity $M_F$ with $h^l_i=1$. The black line corresponds to the values, where $A^l R = 1$ is obtained. Above this line a seesaw like scenario will take place, whereas below this line $A^l R \gg 1$ could hold and thus generate a scenario similar to pseudo Dirac neutrinos.}
\label{fig:AlR}
\end{figure}

In this section, we introduce the field content and general properties of the model. A right handed neutrino  is added to the SM particle content. Since it is not charged under the SM gauge groups it is allowed to propagate in the extra dimension while all SM particles are confined to a (3+1) dimensional subspace, called brane. The analysis assumes that the neutrino experiences only one extra dimension, which is not necessarily the case for gravity. \footnote{This can be realized by embedding the SM 3-brane into a 4-brane which itself is embedded into a $3+n$ dimensional space. The right handed neutrino is confined to the 4-brane while gravity feels the entire $4+n$ dimensional spacetime. The realization of such scenarios is discussed e.g. in \cite{Accomando:1999sj} or \cite{Donini:1999px}} \\
The 5-dimensional bulk neutrino and the SM lepton fields are described by:
\begin{align}
 N \left( x^{\mu} , y \right) = \left( \begin{array}{c}
                                       \Psi_1 \left( x^{\mu} , y \right) \\
                                       \bar{\Psi}_2 \left( x^{\mu} , y \right)
                                       \end{array} \right), \quad \quad L \left( x \right) = \begin{pmatrix}
                                       \nu_l \left( x^{\mu} \right) \\
                                       l\left( x^{\mu} \right)
                                       \end{pmatrix}, \quad \quad l_R \left( x^{\mu} \right).
\end{align}
$L\left( x \right)$ and $l_R \left( x \right)$ are the SM lepton fields with $l=e, \mu, \tau$. The $x^{\mu}$, with $\mu$ running from $0$ to $3$, are the usual coordinates, $y$ is the extra dimensional coordinate and $\Psi_1$ and $\Psi_2$ are 5-dimensional two component spinors.~\footnote{Here the notation, $\bar{\psi}_2$ for a particle transforming under the $(0,\frac{1}{2})$ representation of the Lorentz algebra is chosen in analogy to earlier works on this model. One might be more familiar with the $\psi^{\dagger}_2$ notation that is used in \cite{DREINER20101} which is a useful reference for the two component spinor notation.} The extra dimension is compactified on a $S_1 / \mathbb{Z}_2$ orbifold. $\Psi_1$ is chosen to be even and $\Psi_2$ to be odd under a $y \rightarrow -y$ transformation. \\
The SM fields, including the left handed neutrinos, are restricted to a brane at $y=a$. In order to secure $\mathbb{Z}_2$ invariance, it is necessary to introduce another brane at $y=2 \pi R - a$, which is not relevant for the problem and therefore is not mentioned further in the following. For previous discussions of extra dimensional models with branes shifted away from the 
orbifold fixed points compare \cite{Dienes:1998sb,Bhattacharyya:2002vf,Pilaftsis:1999jk,Ioannisian:1999cw}, while in \cite{Antoniadis:1998ig} the first string realization of low scale gravity and braneworlds was given.
 \\
The Lagrangian of the model is given by \cite{Dienes:1998sb,Pilaftsis:1999jk}:
\begin{align}
\mathcal{L} =  \int\limits_0^{2 \pi R} dy \left\lbrace \bar{N} \left(i \gamma^{\mu} \partial_{\mu} + \gamma^5 \partial_y \right) N 
-\frac{M}{2} \left( N^T C^{(5)-1}N + h.c. \right) \right. \nonumber \\ \left. +   
\delta \left( y-a \right) \left[ \frac{h_1^l}{M_F^{1 /2}} L \tilde{\Phi}^* \Psi_1 + \frac{h_2^l}{M_F^{1 /2}} L \tilde{\Phi}^* \Psi_2 \right] + \delta \left( y-a \right) \mathcal{L}_{SM} \right\rbrace.
\label{eq:L}
\end{align}
Here $\tilde{\Phi} = i \sigma_2 \Phi^{*}$ is the hypercharge conjugate of the SM Higgs doublet $\Phi$ and $\mathcal{L}_{SM}$ is the SM Lagrangian.
The 5D $\gamma$ matrices and the charge conjugation operator are defined as \cite{Pilaftsis:1999jk}:
\begin{align}
 \gamma^{\mu}=\begin{pmatrix}
                     0 &\sigma^{\mu} \\
                     \bar{\sigma}^{\mu}  &0
              \end{pmatrix} \quad \quad
 \gamma^5 =   \begin{pmatrix}
                     -1_2 &0 \\
                     0  &1_2
              \end{pmatrix} \quad \quad C^{5} =- \gamma_1 \gamma_3 = \begin{pmatrix}
                     -i \sigma_2 &0 \\
                     0  &-i \sigma_2
              \end{pmatrix}, \nonumber 
 \end{align}
where $\sigma^{\mu}=(1_2,\sigma)$ and $\bar{\sigma}^{\mu}=(1_2,-\sigma) $ with $\sigma$ being the usual 4D Pauli matrices and $C^{5}$ is the 5-dimensional analog to charge conjugation in 4 dimensions while, as discussed in \cite{Pilaftsis:1999jk}, the gauge invariant mass term $N^T C^{(5)-1}N$ is not a true Majorana mass term. However, after integrating out the extra dimension a Majorana mass term in the effective 4-dimensional theory is obtained. 
The fundamental dimensionless 5D Yukawa couplings are defined as $h_{1/2}^l$ and $M_F$ is the fundamental higher dimensional scale of gravity. \\ 
In a further step, it is necessary to perform the y-integration in the Lagrangian \ref{eq:L}.
The fields $\Psi_1$ and $\Psi_2$ are symmetric and antisymmetric under the $y$ to $-y$ transformation. Consequently, they can be expanded in a Fourier series:
\begin{align}
 \Psi_1 \left( x^{\mu}, y \right) &= \frac{1}{\sqrt{2 \pi R}} S_0 \left( x^{\mu} \right) + \frac{1}{\sqrt{\pi R}} \sum\limits_{k=1}^{\infty} S_k \left( x^{\mu} \right) \cos \left( \frac{ky}{R} \right) \\
 \Psi_2 \left( x^{\mu}, y \right) &= \frac{1}{\sqrt{\pi R}} \sum\limits_{k=1}^{\infty} A_k \left( x^{\mu} \right) \sin \left( \frac{ky}{R} \right).
\end{align}
In the next step, the series expansion for $\Psi_1$ and $\Psi_2$ is substituted into the Lagrangian \eqref{eq:L} and the $y$-Integration is performed, yielding the following effective Lagrangian:
\begin{align}
\mathcal{L}_{eff} = \mathcal{L}_{SM} + \bar{S}_0(i \bar{\sigma}_{\mu} \partial^{\mu})S_0 + \left(h_1^{l(0)} L \tilde{\Phi}^* S_0 - \frac{M}{2} S_0 S_0 + h.c. \right) \\
+ \sum\limits_{k=1}^{\infty} \left[ \bar{S}_k(i \bar{\sigma}_{\mu} \partial^{\mu})S_k + \bar{A}_k(i \bar{\sigma}_{\mu} \partial^{\mu})A_k + \frac{k}{R} \left(\bar{A}_k \bar{S}_k+ S_k A_k \right) \nonumber \right. \\ \left.
- \frac{M}{2} \left( S_k S_k+ \bar{A}_k \bar{A}_k + h.c. \right) + \sqrt{2} \left( h_1^{l(k)} L \tilde{\Phi}^* S_k + h_2^{l(k)} L \tilde{\Phi}^* A_k + h.c.  \right) \right]. \nonumber
\end{align}
The $\delta$ function in the Yukawa coupling terms leads to 4D Yukawa couplings $ h_1^{l(k)}$ and $ h_2^{l(k)}$, depending on the brane shift $a$ away from the fixed points:
\begin{align}
 h_1^{l(k)} = \frac{h_1^l}{\left(2 \pi M_F R \right)^{\frac{1}{2}}} \cos \left( \frac{ka}{R} \right), \quad h_2^{l(k)} = \frac{h_2^l}{\left(2 \pi M_F R \right)^{\frac{1}{2}}} \sin \left( \frac{ka}{R} \right). \label{eq:yukawa1}
\end{align}
Note that $h_2^{l(k)}$ vanishes for $a=0$ due to the $\mathbb{Z}_2$ invariance and the fact that $\Psi_2$ is odd under $y \rightarrow -y$. Using the relation between the fundamental scale of gravity $M_F$ and the Planck scale $M_P$ in dependence of the number of extra dimension $n$ assuming extra dimensions with an equal radius $R$, $M_P = \left( 2 \pi M_F R \right)^{\frac{n}{2}} M_F$, it is obtained: 
\begin{align}
 h_1^{l(k)} = \left( \frac{M_F}{M_P} \right)^{\frac{1}{n}} h_1^l \cos \left( \frac{ka}{R} \right), \quad h_2^{l(k)} = \left( \frac{M_F}{M_P} \right)^{\frac{1}{n}}  h_2^l \sin \left( \frac{ka}{R} \right).
\end{align}
Thus, the 5D Yukawa couplings, expected to be of $\mathcal{O} \left( 1 \right)$, are suppressed by $\bar{h}_i^l = \left( \frac{M_F}{M_P} \right)^{\frac{1}{n}} h_i^l$.  \\
Rewriting the fields $S$ and $A$ into a new basis, the so called weak basis for Kaluza Klein Weyl Spinors, yields:
\begin{align}
\chi_{\pm k} = \frac{1}{\sqrt{2}} \left( S_k \pm A_k \right).
\end{align} 
This leads to the following kinetic term in the Lagrangian (for a more detailed calculation see \cite{Dienes:1998sb,Bhattacharyya:2002vf,Pilaftsis:1999jk} that use the same setup):
\begin{align}
\mathcal{L} = \bar{\chi} i \bar{\sigma}^{\mu} \partial_{\mu} \chi - \left( \frac{1}{2} \chi^{T} \mathcal{M} \chi + h.c. \right),
\end{align}
where
\begin{align}
\mathcal{M} = \begin{pmatrix}
      0 & 0 & 0 & m_0^e & m_{+1}^{e} & m_{-1}^{e} & \cdots \\
      0 & 0 & 0 & m_0^{\mu} & m_{+1}^{\mu} & m_{-1}^{\mu} & \cdots \\
      0 & 0 & 0 & m_0^{\tau} & m_{+1}^{\tau} & m_{-1}^{\tau} & \cdots \\
      m_0^{e} & m_0^{\mu} & m_0^{\tau} & M & 0 & 0 & \cdots \\
      m_{+1}^{e} & m_{+1}^{\mu} & m_{+1}^{\tau} & 0 & M+\frac{1}{R} & 0 & \cdots \\
      m_{-1}^{e} & m_{-1}^{\mu} & m_{-1}^{\tau} & 0 & 0 & M-\frac{1}{R} & \cdots \\
      \vdots & \vdots & \vdots & \vdots & \vdots & \vdots & \ddots
     \end{pmatrix} = \begin{pmatrix}
      0 & Y^T \\
      Y & \mathcal{M}_{\text{KK}}
     \end{pmatrix} 
     \label{eq:MMatrix}
\end{align}
and $\chi^{T} = (\nu_l^{e},\nu_l^{\mu},\nu_l^{\tau}, \chi_0, \chi_{+1}, \chi_{-1}, \ldots)$ hold. The $m_k$ are a combination of the 
Yukawas $h_1^{(k)}$ and $h_2^{(k)}$:
\begin{align}
m_k^l = \frac{v}{\sqrt{2}} \left[ \bar{h}_1^l \cos \left( \frac{ka}{R} \right) + \bar{h}_2^l  \sin \left( \frac{ka}{R} \right) \right] = A^l \cos
\left( \frac{ka}{R} + \Phi^l \right).
\end{align}
where $A^l = \frac{v}{\sqrt{2}} \sqrt{ \left( \bar{h}_1^l \right)^2 + \left( \bar{h}_2^l \right)^2}$, $\Phi^l = -\arctan \left( \frac{h_2^l}{h_1^l} \right)$ and $v$ is the VEV of the Higgs. \\
Now the $\chi_k$ are rearranged in a way, that $\chi_0$ corresponds to the smallest diagonal entry $ |M_0| = \text{min}|M \pm \frac{k}{R} |$ 
in the mass matrix \cite{Dienes:1998sb}, with $|M_0|<\frac{1}{2R}$. Therefore, the mass scale $M$ is irrelevant for the neutrino masses and replaced by $R^{-1}$. Assuming the minimum to lie at $k=k_0$ the phases in the $m_k^l$ need to be changed:
\begin{align}
\Phi^l = -\arctan \left( \frac{h_2^l}{h_1^l} \right) - \frac{k_0 a}{R}.
\end{align}
Then, the four component spinor vector $\Psi_{\nu}$ is defined:
\begin{align}
\Psi_{\nu}^T = \left( \begin{pmatrix} \nu_l \\ \bar{\nu}_l \end{pmatrix} , \begin{pmatrix} \chi_{k_0} \\ \bar{\chi}_{k_0} \end{pmatrix}, 
 \begin{pmatrix} \chi_{k_0+1} \\ \bar{\chi}_{k_0+1} \end{pmatrix}, \begin{pmatrix} \chi_{k_0-1} \\ \bar{\chi}_{k_0-1} \end{pmatrix} , \cdots \right) . 
\end{align}
Hence, the kinetic term can be written as: 
\begin{align}
\mathcal{L}_{kin} = \frac{1}{2} \bar{\Psi}_{\nu} \left( i \slashed{\partial} - \mathcal{M}  \right) \Psi_{\nu}.
\end{align}
At this point, the dimensionless product $A^l R$ has to be analyzed since only for the case that $ A^l R \ll 1$ holds a seesaw kind of behavior is possible. By using equation \eqref{eq:yukawa1}, the relation between the new scale of gravity $M_F$ and the inverse Radius $R^{-1}$ 
\begin{align}
R^{-1} = 2 \pi \left( \frac{M_F}{M_P} \right)^{\frac{2}{n}} M_F \, \label{eq:RMF} ,
\end{align}
and assuming $h^l_i$ to be of $\mathcal{O} \left( 1 \right)$, we obtain:
\begin{align}
A^l R = \left( \frac{M_P}{M_F} \right)^{\frac{n+1}{n}} \frac{v}{M_P} \frac{1}{2 \pi}. 
\end{align}
The value of $\log_{10} \left( A^l R \right)$ is shown in figure \ref{fig:AlR}. For a small $\frac{M_F}{M_P}$, which is necessary to solve the hierarchy problem, a larger number of extra dimensions is needed to obtain a seesaw kind of behavior. Note that in principle even in the red regions of the plot it is possible to obtain a seesaw kind of behavior by choosing a small value for the 5D Yukawa coupling. 

\section{Neutrino Masses and Mixing}\label{sec:mass+pmns}
To obtain the neutrino masses the eigenvalues of \eqref{eq:MMatrix} have to be found. Calculating the characteristic polynomial results in:
\begin{align}
0=P \left[K_3 \lambda^3 +K_2 \lambda^2 + K_1 \lambda + K_0  \right],
\label{eq:CP}
\end{align}
where 
\begin{align}
K_3 &= 1 \\
K_2 &=\sum \limits_{k=-\infty}^{\infty} \sum \limits_{F} \frac{\left( m_k^F \right)^2}{M_0+\frac{k}{R}-\lambda} \\
K_1 &= \sum \limits_{k,j=-\infty}^{\infty} \sum \limits_{F_1>F_2} \frac{\left( m_k^{F_1} \right)^2 \left( m_j^{F_2} \right)^2 - m_k^{F_1} m_j^{F_1} m_k^{F_2} m_j^{F_2} }{\left( M_0+\frac{k}{R}-\lambda \right) \left( M_0+\frac{j}{R}-\lambda \right)} \\
K_0 &= \sum \limits_{k,j,l=-\infty}^{\infty} \sum \limits_{F_1,F_2,F_3}  \frac{- m_k^e m_j^{\mu} m_l^{\tau} \varepsilon_{F_1 F_2 F_3} m_k^{F_1} m_j^{F_2} m_l^{F_3}}{\left( M_0+\frac{k}{R}-\lambda \right) \left( M_0+\frac{j}{R}-\lambda \right) \left( M_0+\frac{l}{R}-\lambda \right)} \\
P &= \prod \limits_{k=-\infty}^{\infty} M_0 + \frac{k}{R} - \lambda .
\end{align}
The sums over $F$ run over the flavors and $F_1>F_2$ has to be understood according to the mass ordering of the charged $SU(2)$ partners. Firstly, we find that $M_0 \pm \frac{k}{R}$ is not a solution of the equation, since the term in the brackets of eq. \eqref{eq:CP} is divergent for $\lambda \rightarrow M_0 \pm \frac{k}{R}$. \\
Secondly, if the $m_k^F$ \eqref{eq:yukawa1} factorize into a $k$ and into a $F$ dependent part, $m_k^F=m_k m^F$, the factors $K_0$ and $K_1$ are vanishing, resulting in two mass eigenvalues equal to zero. Since for $A^l R \ll 1$ the three lightest eigenvalues should correspond to the three active neutrinos, this would lead to only one mass difference. \\
The $m_k^F$ are factorizable if $\Phi^e = \Phi^{\mu} = \Phi^{\tau}$ and/or $a=0,\frac{\pi R}{2},\pi R$. Consequently, it is not possible to generate two mass differences without a brane shift away from the orbifold fixed points, which means $a \neq 0, \pi R$. Additionally, $a=\frac{\pi R}{2}$ is also forbidden, since this localization of the brane leads to a vanishing contribution of $\Psi_1$ instead of a vanishing contribution of $\Psi_2$ as for $a=0,\pi R$, resulting in a factorizable $m_k^F$. \\ 
Next, the infinite sums in the $K_i$ are solved. This is done explicitly in appendix \ref{sec:appA} and the following result is obtained:
\begin{align}
  \frac{S \left( F_1 , F_2, \lambda \right)}{\pi R A^{F_1} A^{F_2}} =  &\left[ \cot \left(  \pi R \left[M_0 - \lambda \right] \right) \cos \left( \Phi^{F_1} - a \left[M_0 - \lambda \right] \right) \cos \left( \Phi^{F_2} - a \left[M_0 - \lambda \right] \right) \right. \nonumber \\ &- \left.\frac{1}{2} \sin \left( \Phi^{F_1} + \Phi^{F_2} - 2a \left[M_0 - \lambda \right] \right) \right]. \label{eq:sum1}
\end{align}  
Therewith, the coefficients take the form:
\begin{align}
K_2 &= \sum \limits_{F} S\left( F , F, \lambda \right) 
&K_1 &= \sum \limits_{F_1>F_2} S \left( F_1 , F_1, \lambda \right) S \left( F_2 , F_2, \lambda \right) - S \left( F_1 , F_2, \lambda \right)^2 \nonumber \\
K_3 &= 1 &K_0 &= - \sum \limits_{F_1,F_2,F_3} \varepsilon_{F_1 F_2 F_3} S \left( \text{e} , F_1, \lambda \right) S \left( \mathrm{\mu} , F_2, \lambda \right) S \left( \mathrm{\tau} , F_3, \lambda \right)=0. \nonumber
\end{align}
Since $K_0=0$, one eigenvalue is always zero, meaning the lightest neutrino mass eigenvalue vanishes.
\subsection{Neutrino Masses for $A^F R \ll 1$ }
\begin{figure}
\centering
\includegraphics[width=0.6\textwidth]{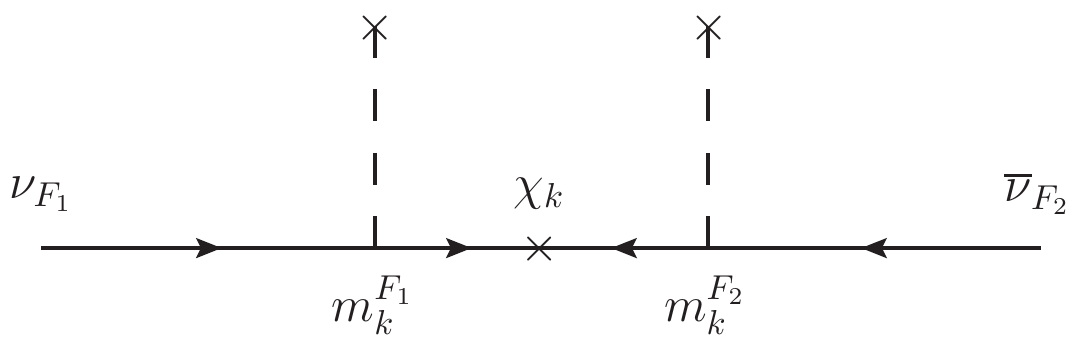}
\caption{Tree level diagram to generate the entries of a effective $3 \times 3$ mass matrix}
\label{fig:feynman1}
\end{figure}

In the following, the neutrino mass generation for $A^l R \ll 1$ is discussed in more detail. \footnote{The opposite case $A^l R \gg 1$ is more similar to pseudo Dirac Neutrinos. Nevertheless, there is a major difference, since for some $k$ the masses of the KK excitations, $\pm k R^{-1}$, become larger than the Dirac Masses $A^l$. In contrast to the considerations for $A^l R \ll 1$, where three mostly left handed neutrinos are obtained, for this scenario a large number of $\left( A^l R \right)^2$ neutrino mass eigenstates with an $\frac{1}{10} \left( A^l R \right)^{-2}$ fraction being left handed is generated. All other eigenstates have a significantly lower left handed contribution. A quick calculation shows that the mass eigenstates are almost equidistant separated by $R^{-1}$. At this point one could study whether it is possible to explain the observed neutrino oscillation phenomena with such a large number of neutrino states with nearly the same left handed part and almost equal mass differences. However, this is not further discussed here. }
We are mostly interested in the masses of the active neutrinos and to obtain analytic expressions for them. For that reason, it is assumed that the three lowest eigenvalues correspond to the three active neutrino masses. Consequently, they should be found by performing a series expansion for $\lambda$ around zero up to third order in equation \eqref{eq:CP}. The expansion results in $\sum \limits_{i=0}^3 C_i \lambda^i = 0$ with:
\begin{align}
 C_1 &= \frac{\pi^2 R^2}{8} \left[ \sum \limits_{F_1 > F_2} \left( A^{F_1} \right)^2 \left( A^{F_2} \right)^2 \left( \cos \left( 2 \left[ \Phi^{F_1} - \Phi^{F_2} \right] \right) - 1 \right) \right] \nonumber \\
 C_2 &= \frac{\pi R}{2 \sin \left( M_0 \pi R \right)} \left[ \sum \limits_F \left( A^{F} \right)^2 \left( \cos \left( M_0 \pi R \right) + \cos \left(  M_0 \left[ 2a - \pi R \right] - 2 \Phi^F   \right) \right) \right] \nonumber \\
 C_3 &= \frac{1}{2 \sin \left( M_0 \pi R \right)^2} \left[ 1 - \cos \left( 2 M_0 \pi R \right) + \sum \limits_F  \left( A^F \right)^2 \left( \pi^2 R^2 + \right. \right. \nonumber \\ &\left. \left. \left( \pi R -a \right) \pi R  \cos \left( 2 M_0 a - 2 \Phi^F \right) + a \pi R \cos \left( 2 M_0 \pi R - 2 \Phi^F - 2 M_0 a \right) \right)   \right]. \label{eq:coefficents}
\end{align}
In a first approach, it is assumed that the $\Phi^F$ are all equal. As shown before, equal $\Phi^F$ are leading to two zero mass eigenvalues. The remaining nonzero eigenvalue is calculated to show the behavior of the neutrino mass for different regions of the dimensionless parameter $M_0 \pi R$. Later, the second mass difference is generated by small differences in the $\Phi^F$, $\delta \Phi$. \\
By setting $\Phi^e=\Phi^{\mu}=\Phi^{\tau}= \Phi$, the nonzero eigenvalue results in $\lambda_3 = - \frac{C_2}{C_3}$:
\begin{itemize}
\item $A^F R \ll 1$ \\
The eigenvalue results in:
\begin{align}
\lambda_3 &\approx  \pi R \sum \limits_F \left( A^F \right)^2 \frac{\cos \left( a M_0 - \Phi \right) \cos \left( a M_0 - M_0 \pi R - \Phi \right)}{\sin \left( M_0 \pi R \right)} \nonumber  \\
&= \pi R \sum \limits_F \left( A^F \right)^2 f \left( a, M_0, R, \Phi \right). \label{eq:limitseesaw}
\end{align}
The result splits into two products. The first one $\pi R \sum \limits_F \left( A^F \right)^2$ is similar to the well known seesaw mass term ($\frac{m^2}{M}$). The mass of the heavy right handed neutrino is replaced by $R^{-1}$. The mass of the introduced right handed bulk neutrino no longer has to be very large, instead a small extra dimension in comparison to the $A^F$ is required. \\
The second factor is a function $f \left( a, M_0, R, \Phi \right)$ of the 'form' of the extra dimension described by the placement of the brane in the extra dimension $a$, the lowest diagonal entry in the mass matrix for the KK states $M_0$, the radius of the extra dimension $R$ and the phase $\Phi$. This allows to lower the neutrino mass by the function $f$. \\
Furthermore, if we assume the 5D Yukawa couplings to be of $\mathcal{O} \left( 1 \right)$ and substitute eq. \eqref{eq:RMF} and eq. \eqref{eq:yukawa1} for $R$ and $A_F$, respectively, the product $\pi R \sum_F \left( A^F \right)^2$ yields:
\begin{align}
\pi R \sum_F \left( A^F \right)^2 \sim \frac{v^2}{M_F}. \label{eq:nuscale}
\end{align} 
Here, $v$ is the Higgs VEV and $M_F$ is the new fundamental scale of gravity. Thus, the first factor in $\lambda_3$ can be interpret as the typical type I seesaw formula with $M_F$ playing the roll of the heavy right handed neutrino mass. However, if $M_F$ is of $\mathcal{O} \left(10 \, \mathrm{TeV} \right)$, the first factor in $\lambda_3$ is of $\mathcal{O} \left( 1 \, \mathrm{GeV} \right)$. Consequently, the second factor $f \left( a , M_0 , R, \Phi \right)$ is required to be small in order to achieve a neutrino mass of $\mathcal{O}\left( 10^{-2} \, \mathrm{eV} \right)$.  \\ 
Another possibility to realize $m_\nu \sim 10^{-2} \, \mathrm{eV}$ is to allow for larger scales $M_F$. Within this setup the correct neutrino mass could also be obtained with a larger $f \left( a , M_0 , R, \Phi \right)$. Moreover, for $M_F \geq 10^{11} \, \mathrm{GeV}$ a seesaw like scenario (compare with figure \ref{fig:AlR}) can be realized within a symmetric setup, i.e. gravity is propagating in same number of extra dimensions as the right handed neutrino does.  \\
It should be noticed that expression \eqref{eq:limitseesaw} in the limit of $a \rightarrow 0$ does not coincide with the result obtained for $A^F R \ll 1$ with a vanishing brane shift, 
\begin{align}
\lambda_3 = -\pi R \cot \left( \pi M_0 R \right) \sum \limits_F \left( \bar{h}_1^F \right)^2.
\end{align} 
This issue can be resolved by assuming that new physics enters above the scale $M_F$, leading to an exponential suppression of KK-excitations with masses greater than $M_F$. For a more detailed discussion see chapter 4 of \cite{Bhattacharyya:2002vf}. However, the presented formula for the eigenvalue is valid as long as $a \gg M_F^{-1}$ holds. If in the following a small $a$ is considered, it is important to keep in mind that $a \gg M_F^{-1}$ still holds.
\item $M_0 \pi R \rightarrow 0 \quad \text{and} \quad M_0 \gg A^F \Rightarrow \lambda_3=-\left( A^F \right)^2 M_0^{-1} \cos \left( \Phi \right)^2$ \\
Here the assumption $M_0 \ll R^{-1}$ is added. Thus, the important scale for the seesaw mechanism is $M_0$ instead of $R^{-1}$. Within this limit, only the lightest KK excitation is relevant for the neutrino mass generation.
\end{itemize}
Another advantage of the limit $A^F R \ll 1$ is that the KK excitation can be integrated out. As a consequence, it is possible to obtain an effective three by three mass matrix for the active neutrinos by calculating the diagrams presented in figure \ref{fig:feynman1}. 
\begin{align}
&\mathcal{M}_{F_1,F_2}^{\text{eff}}=\sum \limits_{k=-\infty}^{\infty} \frac{m_k^{F_1} m_k^{F_2}}{M_0 + \frac{k}{R}} = Y^T \mathcal{M}_{\text{KK}}^{-1} Y= S \left( F_1, F_2 , 0 \right), \label{eq:effMM}
\end{align}
where $S$ is the solution of the infinite sum \eqref{eq:sum1}. Calculating the eigenvalues of $\mathcal{M}_{F_1,F_2}$ with equal $\Phi^F$ yields the same eigenvalue as presented in the approximation $A^F R \ll 1$ \eqref{eq:limitseesaw}. \\
The next step is to analyze the influence of slightly different $\Phi^F$. For that it is defined:
\begin{align}
 \Phi^e=\Phi \quad \quad \Phi^{\mu}=\Phi + \delta \Phi \quad \quad \Phi^{\tau}=\Phi + r \delta \Phi
\end{align}
To simplify the expressions for the neutrino masses, a series expansion in $\delta \Phi$ up to leading order is performed. The expansion results in \eqref{eq:limitseesaw} for $\lambda_3$ and in
\begin{align}
 \lambda_2 = \frac{\pi R \sin \left( M_0 \pi R \right) \left[ \left( A^e A^{\mu} \right)^2 + \left( A^e A^{\tau} \right)^2 r^2 + \left( A^{\mu} A^{\tau} \right)^2 \left( r-1 \right)^2 \right]}{4 \left[ \cos \left( a M_0 - \Phi \right) \cos \left( a M_0 - M_0 \pi R - \Phi \right) \right] \left[ \left( A^e \right)^2 + \left( A^{\mu} \right)^2 + \left( A^{\tau} \right)^2 \right] } \delta \Phi^2.
\end{align}
Moreover, we define: $A^F=c_F Y$, with $c_e=1$, and 
\begin{align}
s \left( c_{\mu}, c_{\tau} \right) &= 1 + c_{\mu}^2 + c_{\tau}^2 \\
w \left( c_{\mu}, c_{\tau} , r \right) &= c_{\mu}^2 + c_{\tau}^2 r^2 + c_{\mu}^2 c_{\tau}^2 \left( r-1 \right)^2 \, .
\end{align}
With these definitions, the eigenvalues of $\mathcal{M}_{F_1,F_2}^{\text{eff}}$ are given by:
\begin{align}
\lambda_1&= 0 \\
\lambda_2&= -\frac{\pi R Y^2}{4} \frac{w \left( c_{\mu}, c_{\tau} , r \right)}{s \left( c_{\mu}, c_{\tau} \right) f \left( a , M_0 , R, \Phi  \right)} \delta \Phi^2 \label{eq:l+} \\ 
\lambda_3&= \pi R Y^2 s \left( c_{\mu}, c_{\tau} \right) f \left( a , M_0 , R, \Phi  \right). \label{eq:l-}
\end{align}
Eventually, we want to comment on current collider bounds on large extra dimensions \cite{Aad:2015zva}. The ATLAS collaboration found an lower limit on the fundamental scale of gravity $M_F$ of $ \frac{M_F}{TeV} \geq \left( 5.25 , 4.11 , 3.57, 3.27, 3.06  \right)$ for $n=\left( 2, 3, 4, 5, 6 \right)$ extra dimensions. These limits can be translated into upper bounds on the radius of the extra dimension by applying formula \eqref{eq:RMF}. The limits are compatible with the observed neutrino masses within the presented framework. The correct neutrino mass scale can be achieved by either choosing a small $R$ (corresponding to a larger $M_F$) or a small $\delta \phi$ since $\lambda_2 \lambda_3 \sim \pi R Y^2 \delta \phi^2$, while the correct ratio for the eigenvalues can be accommodated for by choosing a suitable $f \left( a , M_0 , R, \Phi  \right)$ since $\frac{\lambda_2}{\lambda_3} \sim \left( \frac{\delta \phi}{f \left( a , M_0 , R, \Phi  \right)} \right)^2$. 
\subsection{Neutrino Mixing in Leading Order in $\delta \Phi$ for $A^l R \ll 1$}
In the following considerations only the case $A^l R \ll 1$ is considered, which was capable of generating a seesaw like scenario. Furthermore, it is also possible to obtain a good approximation for the mixing matrix by diagonalizing $\mathcal{M}_{F_1,F_2}^{\text{eff}}$ \ref{eq:effMM}, by $U^T \mathcal{M}_{F_1,F_2}^{\text{eff}} U = \mathcal{M}_{\text{diag}}$. The obtained $U$ will be unitary, while the exact three by three PMNS matrix is not. The deviation from unitary, $\sim Y^T \mathcal{M}_{\text{KK}}^{-2} Y $, is analyzed in more detail in section \ref{sec:UVandPheno}. \\
Calculating the entries of the mixing matrix $U$ up to leading order in $\delta \Phi$ with the assumption of a normal mass hierarchy $|\lambda_3|>|\lambda_2|$, yields:
\begin{align}
U = \begin{pmatrix}
\frac{c_{\mu} c_{\tau} \left( r -1 \right)}{\sqrt{w \left( c_{\mu} , c_{\tau}, r \right)}} & \frac{c_{\mu}^2 + r c_{\tau}^2}{\sqrt{s \left( c_{\mu} , c_{\tau} \right) w \left( c_{\mu}, c_{\tau} , r \right)}} & \frac{1}{\sqrt{s \left( c_{\mu} , c_{\tau} \right)}} \\
-\frac{c_{\tau} r}{\sqrt{w \left( c_{\mu} , c_{\tau}, r \right)}}  & \frac{c_{\mu} c_{\tau}^2 \left( r - 1 \right) - c_{\mu}^2}{\sqrt{s \left( c_{\mu} , c_{\tau} \right) w \left( c_{\mu}, c_{\tau} , r \right)}} & \frac{c_{\mu}}{\sqrt{s \left( c_{\mu} , c_{\tau} \right)}} \\
\frac{c_{\mu}}{\sqrt{w \left( c_{\mu} , c_{\tau}, r \right)}} & -\frac{c_{\tau} \left( c_{\mu}^2 \left( r - 1 \right) + r \right)}{\sqrt{s \left( c_{\mu} , c_{\tau} \right) w \left( c_{\mu}, c_{\tau} , r \right)}} & \frac{c_{\tau}}{\sqrt{s \left( c_{\mu} , c_{\tau} \right)}} \\
\end{pmatrix}. \label{eq:UPMNS}
\end{align}
Every entry contains a zeroth order contribution in $\delta \Phi$. Remarkably, this approximated result only depends on three parameters of the model: $c_{\mu}$, $c_{\tau}$ and $r$. Thus, comparing this form of $U$ with the standard parametrization of the neutrino mixing matrix excluding the Majorana phases, which are irrelevant for neutrino oscillations,
\begin{align}
\begin{pmatrix}
c_{12} c_{13} & s_{12} c_{13} & s_{13} e^{-i \delta} \\
-c_{23} s_{12} - s_{23} s_{13} c_{12} e^{i \delta} & c_{23} c_{12} - s_{23} s_{13} s_{12} e^{i \delta} & s_{23} c_{13} \\
s_{23} s_{12} - c_{23} s_{13} c_{12} e^{i \delta} & -s_{23} c_{12} - c_{23} s_{13} s_{12} e^{i \delta} & c_{23} c_{13}   
\end{pmatrix},
\end{align}
allows to identify these three parameters with the mixing angles and results in a predictive framework. The CP violating phase $\delta$ is zero in our scenario \footnote{Considering complex Yukawa couplings would allow for a nonzero $\delta$. In the light of the hint for a non-vanishing $\delta \approx - \frac{\pi}{2}$, it might be interesting to investigate the influence of a nonzero CP phase on the parameter space of the model and therefore on the LFV observables discussed in chapter \ref{sec:UVandPheno}. For example, in the case of $\delta = - \frac{\pi}{2}$ the ratios given in the equations \eqref{eq:ct} result in $c_\mu^2 = -\sin^2 \left( \Theta_{23} \right) \sqrt{1+ \sin \left( \Theta_{13} \right)^{-2}} $ and $c_\tau^2 = -\cos^2 \left( \Theta_{23} \right) \sqrt{1+ \sin \left( \Theta_{13} \right)^{-2}}$.}, since real Yukawa couplings were assumed. Consequently, $c_{\mu}$, $c_{\tau}$ and $r$ are given by:
\begin{align}
c_{\mu}^2 &= \cot^2 \left( \Theta_{13} \right) \sin^2 \left( \Theta_{23} \right) \quad \quad c_{\tau}^2 = \cot^2 \left( \Theta_{13} \right) \cos^2 \left( \Theta_{23} \right) \label{eq:ct} \\
r &= \frac{\frac{\tan \left( \Theta_{12} \right)}{\sin \left( \Theta_{13} \right)} + \tan \left( \Theta_{23} \right)}{\frac{\tan \left( \Theta_{12} \right)}{\sin \left( \Theta_{13} \right)} - \cot \left( \Theta_{23} \right)} \quad , \quad \text{for $r>1$} \label{eq:r1} \\
r &= \frac{\frac{\tan \left( \Theta_{12} \right)}{\sin \left( \Theta_{13} \right)} - \tan \left( \Theta_{23} \right)}{\frac{\tan \left( \Theta_{12} \right)}{\sin \left( \Theta_{13} \right)} + \cot \left( \Theta_{23} \right)} \quad , \quad \text{for $r<1$} \label{eq:r2}
\end{align}
Present neutrino oscillation data for the mixing angles (see table \ref{tab:massmix}) \cite{Gonzalez-Garcia:2015qrr} is used to obtain regions for the parameters $c_{\mu}$, $c_{\tau}$ and $r$. 
\begin{table}[h]
\centering
\begin{tabular}{|c|c c|c c|}\hline
Param. & NO Best Fit & NO $3 \sigma$ & IO Best Fit & IO $3 \sigma$ \\ \hline
$\sin^2 \left( \Theta_{12} \right)$ & $0.304$ & $0.270 \rightarrow 0.344$ &$0.304$& $0.270 \rightarrow 0.344$ \\
$\sin^2 \left( \Theta_{23} \right)$ & $0.452$ & $0.382 \rightarrow 0.643$ &$0.579$& $0.389 \rightarrow 0.644$ \\
$\sin^2 \left( \Theta_{13} \right)$ & $0.0218$ & $0.0186 \rightarrow 0.0250$ &$0.0219$& $0.0188 \rightarrow 0.0251$ \\ \hline
$\Delta m_{21}^2 / 10^{-5} \, \mathrm{eV}^2$ & $7.50$ & $7.02 \rightarrow 8.09$ & $7.50$ & $7.02 \rightarrow 8.09$ \\
$\Delta m_{31}^2 / 10^{-3} \, \mathrm{eV}^2$ & $2.457$ & $2.317 \rightarrow 2.607$ & $-2.449$ & $-2.590 \rightarrow -2.307$ \\ \hline
\end{tabular}
\caption{Three-flavor oscillation parameters from \cite{Gonzalez-Garcia:2015qrr} } \label{tab:massmix}
\end{table}
%
%
%
The possible values for $r$ are obtained from the equations \eqref{eq:r1} and \eqref{eq:r2} while the values for $c_{\mu}$ and $c_{\tau}$ are obtained from the equations \ref{eq:ct}. \\
%
The ordering $m_1=|\lambda_1|<m_2=|\lambda_2|<m_3=|\lambda_3|$ is not the only possible ordering for the mass eigenvalues $\lambda_i$. There are three other cases left to discuss (two additional cases are already excluded since $\lambda_1=0$ and $m_2^2>m_1^2$ has to be satisfied). The other three cases are:
\begin{itemize}
 \item Case II: $m_1=|\lambda_1|<m_2=|\lambda_3|<m_3=|\lambda_2|$ (NO) 
 \item Case III: $m_1=|\lambda_2|<m_2=|\lambda_3|>m_3=|\lambda_1|$ (IO) 
 \item Case IV: $m_1=|\lambda_3|<m_2=|\lambda_2|>m_3=|\lambda_1|$ (IO) 
\end{itemize}
The procedure to obtain expressions for the parameters is the same as presented for case I and is repeated for the other cases. The possible $3 \sigma$ regions for the parameters are presented in table \ref{tab:par1}. Note that for case II it is not possible to find an analytic expression for the parameters in dependence of the mixing angles. 
\begin{table}
\centering
\begin{tabular}{|c |c c | c c | c c |} \hline
Case & $c_{\mu}^2$ BF & $c_{\mu}^2$ $3 \sigma$ &  $c_{\tau}^2$ BF & $c_{\tau}^2$ $3 \sigma$ &  $r$ BF & $r$ $3 \sigma$ \\ \hline
I&$20.3$ &$14.9 \rightarrow 33.9$ &24.6 &$13.9 \rightarrow 32.6$ & 1.6 &$1.45 \rightarrow 1.80$  \\  
I&$20.3$ &$14.9 \rightarrow 33.9$ &24.6 &$13.9 \rightarrow 32.6$ &0.64 & $0.55 \rightarrow 0.69$ \\ \hline
II& 0.59 &$0.31 \rightarrow 0.74 $& 1.78 & $1.52 \rightarrow 2.28$ & -0.12 & -0.14 $\rightarrow$ -0.04 \\
II& 0.46 &$0.16 \rightarrow 0.62 $ & 1.91 & $ 1.64 \rightarrow 2.48 $ & -0.48 & -0.59 $\rightarrow$ -0.20 \\ 
II& 1.08 & $0.41 \rightarrow 1.33 $ & 1.29 & $ 1.10 \rightarrow 2.09$ & -1.62 &  -1.84 $\rightarrow$ -0.52  \\
II& 1.63 & $0.86 \rightarrow 1.97$ & 0.73 & $0.61 \rightarrow 1.51$ & -1.17 & -1.18  $\rightarrow$ -0.41  \\ \hline
III&1.22 & $0.89 \rightarrow 1.97$&1.14 & $0.55 \rightarrow 1.62$ & 1.22 &$1.18 \rightarrow 1.28$ \\
III&1.22 & $0.89 \rightarrow 1.97$&1.14 & $0.55 \rightarrow 1.62$ & 0.82 & $0.78 \rightarrow 0.85$ \\ \hline
IV& 0.30 & $0.23 \rightarrow 0.45$& 0.17 & $0.07 \rightarrow 0.25$& 1.56 & $1.45 \rightarrow 1.80$ \\
IV& 0.30 & $0.23 \rightarrow 0.45$& 0.17 & $0.07 \rightarrow 0.25$& 0.62 & $0.45 \rightarrow 0.68$ \\ \hline
\end{tabular}
\caption{Allowed Parameter regions which reproduce the observed neutrino mixing for the different possible orderings of the mass eigenvalues. The first to cases correspond to the NO and the last two to the IO. The values are obtained by using the Best Fit values (BF) for the mixing angles and the $3 \sigma$ regions, respectively.} \label{tab:par1}
\end{table}
\section{Unitarity Violation and Lepton Flavor Violation}\label{sec:UVandPheno}
\begin{figure}
\centering
\includegraphics[width=0.6\textwidth]{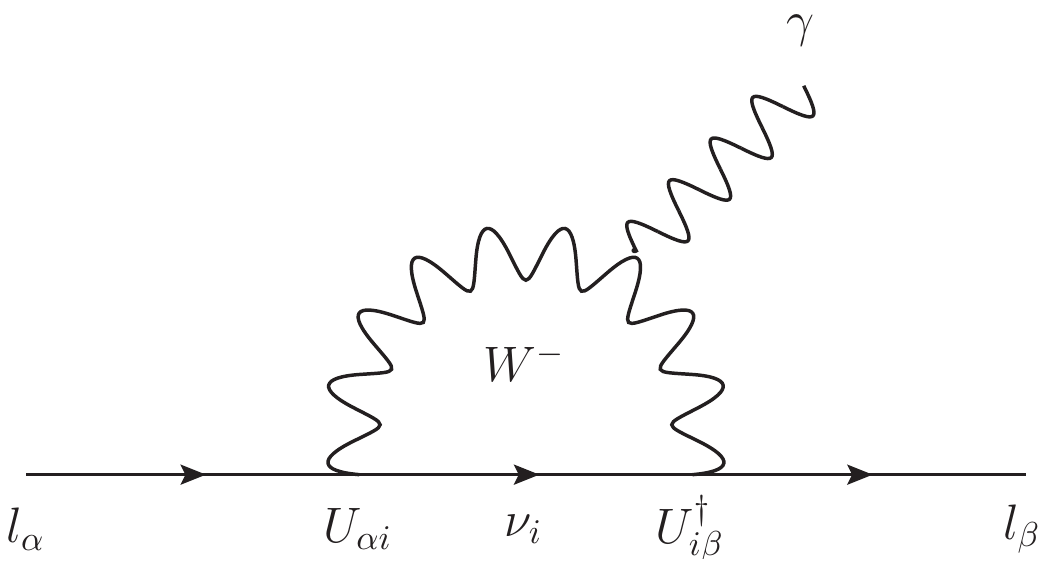}
\caption{Lepton flavor violating decay at one loop}
\label{fig:rarelepton}
\end{figure}

\begin{figure}[h]
\begin{minipage}{0.499\textwidth}
\centering
\includegraphics[width=1\textwidth]{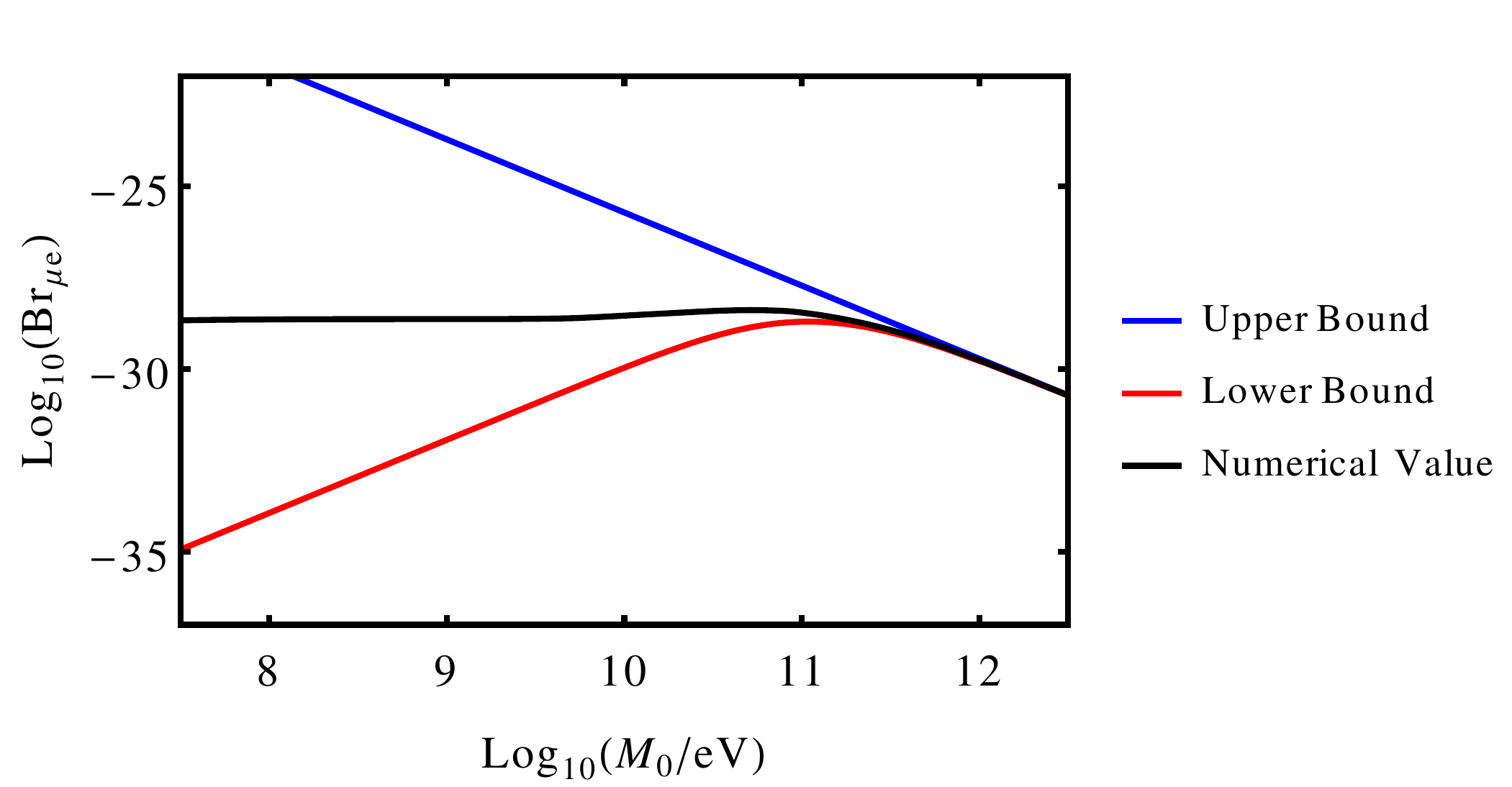}
\end{minipage}
\begin{minipage}{0.499\textwidth}
\centering
\includegraphics[width=1\textwidth]{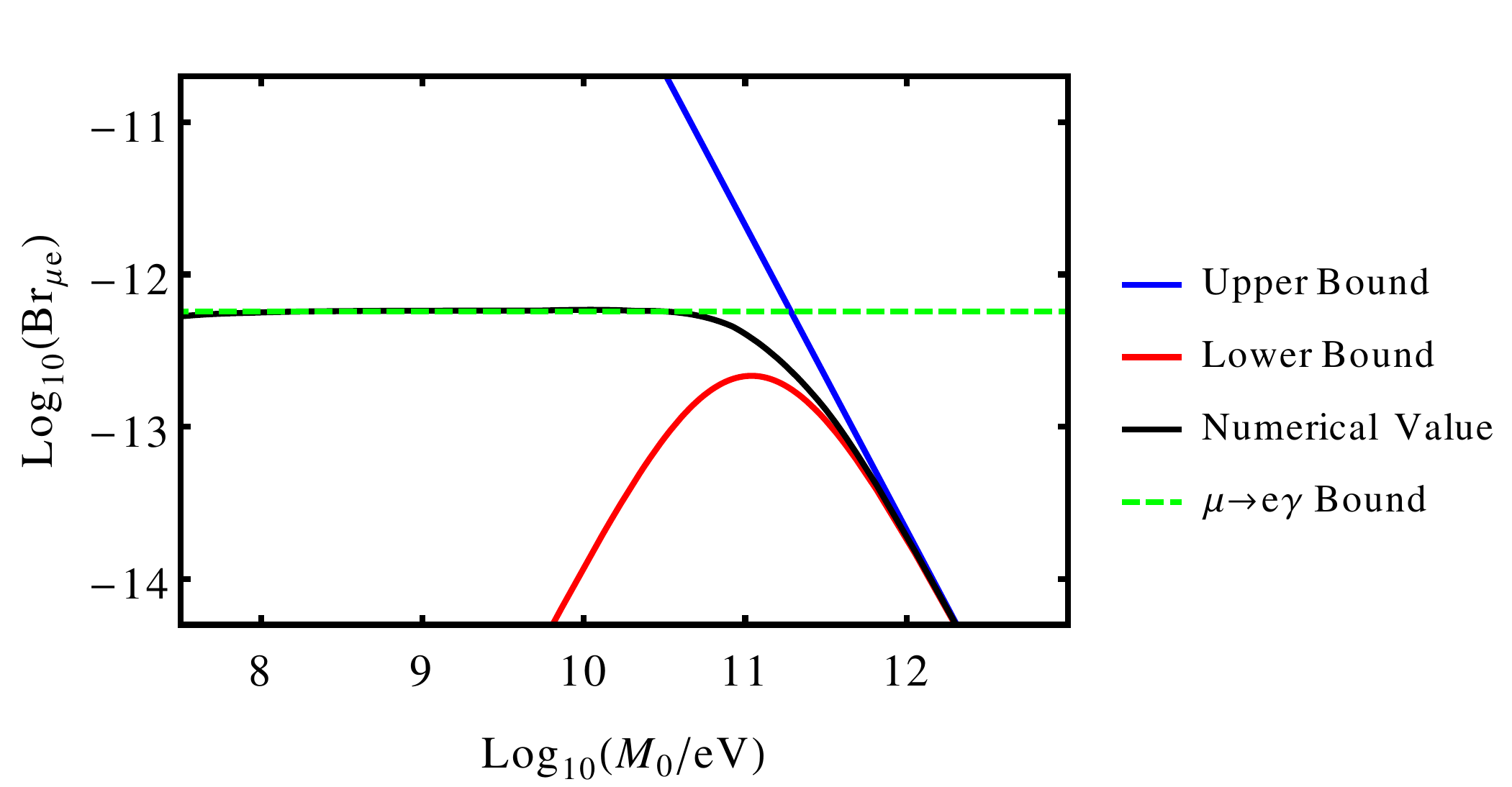}
\end{minipage}
\caption{Upper bound (upper line), lower bound (lower line) and numerically obtained value (intermediate line) for $Br_{\mu e}$ are plotted against $M_0$ for two different configurations of $(q,y,\Phi)$: The left panel shows the branching ratio for $(0.1 , \frac{\pi}{3}, \frac{\pi}{6} )$ and the right panel for $(\epsilon, \frac{\pi}{3}, \frac{\pi}{6})$ with $\epsilon=10^{-9}$. For the configuration in the right panel the branching ratio is close to the experimental bound (dotted line). The numerically estimated value lies always between the two bounds, has a maximum near $M_0 = m_W$ and seems to approach a constant value for $M_0 < m_W$. }
\label{fig:megwoED}
\end{figure}

In the previous section, an approximated non unitary violating mixing matrix for the SM neutrinos was calculated. In this section the unitarity violation of the system is analyzed. 
The deviation of unitarity is given by (calculated in appendix \ref{sec:appB}):
\begin{align}
 \left( \mathcal{E} \right)_{F_1 F_2}=- \left( Y^T \mathcal{M}_{\text{KK}}^{-2} Y \right)_{F_1 F_2} = -\sum \limits_{k=-\infty}^{\infty} \frac{m_k^{F_1} m_k^{F_2}}{\left( M_0 + \frac{k}{R} \right)^2} =: - S_2 \left( F_1, F_2 \right) =\frac{d}{d M_0} S_1 \left(F_1,F_2,0 \right). \label{eq:UV}
\end{align}
Noteworthy is the influence of the unitarity violation on e.g. rare lepton decays or lepton flavor violating $Z$ decays. These influences are discussed in the following. Note that unitarity violation also has an influence on neutrino oscillation. These influences are not further discussed here but e.g. the effect of large extra dimensions on the DUNE experiment is discussed in \cite{Berryman:2016szd}. \\  
As has been pointed out in \cite{Antusch:2014woa,Antusch:2006vwa}, the decay width of rare lepton decays $l_{\alpha} \rightarrow l_{\beta} \gamma$, mediated at one loop level as shown in figure \ref{fig:rarelepton}, strongly depends on the unitarity violation. Furthermore, the ratio of its decay width to the decay width of $l_{\alpha} \rightarrow v_{\alpha} \bar{\nu}_\beta l_{\beta}$ is given by: \cite{Antusch:2014woa,Antusch:2006vwa}
\begin{align}
\frac{\Gamma \left( l_{\alpha} \rightarrow l_{\beta} \gamma \right)}{\Gamma \left( l_{\alpha} \rightarrow l_{\beta} \bar{\nu}_{\beta} \nu_{\alpha} \right)}  = \frac{3 \alpha}{32 \pi} \frac{|\sum_{k=1}^{\infty} U_{\alpha k} U_{k \beta}^{\dagger} F\left( x_k \right)|^2}{\left( U U^{\dagger} \right)_{\alpha \alpha} \left( U U^{\dagger} \right)_{\beta \beta} } .
\end{align}
The matrix $U$ is the mixing matrix as defined in Appendix \ref{sec:appB}. In the sum over $k$, $k=1,2,3$ correspond to the mass eigenvalues of the active neutrinos. The ones corresponding to $k>3$ are the ones close to the masses of the KK excitations. The function $F\left( x_k \right)$ is a loop function with $x_k= \frac{m_{\nu_k}^2}{m_W^2}$, where $m_{\nu_k}$ is the mass of the $k$-th neutrino mass eigenstate and the $F\left( x_k \right)$ is given by:
\begin{align}
 F\left( x \right) = \frac{10 - 43 x+ 78 x^2 -49 x^3 +4 x^4 + 18 x^3 \ln \left( x \right)}{3\left( x-1 \right)^4}
\end{align}
If the sum $\sum \limits_{k=1}^{\infty} U_{\alpha k} U_{k \beta}^{\dagger} F\left( x_k \right)$ is split into $\sum \limits_{k=1}^{3} U_{\alpha k} U_{k \beta}^{\dagger} F\left( x_k \right) + \sum \limits_{k=4}^{\infty} U_{\alpha k} U_{k \beta}^{\dagger} F\left( x_k \right)$ it is reasonable to assume $F \left( x_k \right) \approx \frac{10}{3}$ for $k=1,2,3$, since then $m_{\nu_k} \ll m_W$ holds, what allows to simplify the first sum:
\begin{align}
 \sum \limits_{k=1}^{3} U_{\alpha k} U_{k \beta}^{\dagger} F\left( x_k \right) \approx \frac{10}{3} \left( U_P U_P^T \right)_{\alpha \beta}  \approx \frac{10}{3} \left( \mathcal{E} \right)_{\alpha \beta}.
\end{align}
Since the complete mixing matrix is unitary, $\sum \limits_{k=4}^{\infty} U_{\alpha k} U_{k \beta}^{\dagger}=-\left( \mathcal{E} \right)_{\alpha \beta} $ is valid. For $x \geq 0$ the function $F \left( x \right)$ is always decreasing starting from the value $F\left( 0 \right)=\frac{10}{3}$ and reaching its minimal value for $F \left( \infty \right) = \frac{4}{3}$. Assuming $\mathcal{E}_{\alpha \beta} \ll 1$ and $F\left( x_k \right) = \frac{4}{3}$ for $k \geq 4$, what is equivalent to assuming $M_0 \gg m_W$, allows to find an upper bound for the decay rate or a good approximation for the case $M_0 \gg m_W$, respectively. 
\begin{align}
 \frac{\Gamma \left( l_{\alpha} \rightarrow l_{\beta} \gamma \right)}{\Gamma \left( l_{\alpha} \rightarrow l_{\beta} \bar{\nu}_{\beta} \nu_{\alpha} \right)} \approx \frac{3 \alpha }{32 \pi} |\sum \limits_{k=1}^{3} U_{\alpha k} U_{k \beta}^{\dagger} F\left( x_k \right)+ \sum \limits_{k=4}^{\infty} U_{\alpha k} U_{k \beta}^{\dagger} F\left( x_k \right)|^2 \leq \frac{3 \alpha}{8 \pi} \left( \mathcal{E} \right)_{\alpha \beta}^2
\end{align}
Next, we derive lower bounds on the decay rate for $M_0 \approx m_W$ and $M_0 \ll m_W$. To this end we assume $F \left( x_k \right)=F \left( M_0 \right)$ for $k \geq 4$. This is justified since the Loop function $F \left( \frac{m_i^2}{m_W^2} \right) $ is decreasing with an increasing $m_i$ and the decay rate is 
proportional to $\sum \limits_{i>3} (A- F \left( \frac{m_i^2}{m_W^2} \right) )$. 
Thus, by choosing $m_i = M_0$, which is the lowest KK mass, for all $i$ a lower bound on the decay rate is obtained.
We discuss the following cases:
\begin{itemize}
 \item $M_0 \approx m_W$ \\
 With $F \left( 1 \right)=\frac{17}{6}$, the lower bound results in:
 \begin{align}
  \frac{\Gamma \left( l_{\alpha} \rightarrow l_{\beta} \gamma \right)}{\Gamma \left( l_{\alpha} \rightarrow l_{\beta} \bar{\nu}_{\beta} \nu_{\alpha} \right)} \geq \frac{3 \alpha}{128 \pi} \left( \mathcal{E} \right)_{\alpha \beta}^2 \label{eq:BrmeM0}
 \end{align}
 In comparison with the upper bound, a factor $\frac{1}{16}$ is multiplied to the upper bound.  
\item $M_0 \ll m_W$ \\
A series expansion for small arguments of $F \left( x \right)$ up to first order yields $F  \left( \frac{M_0^2}{m_W^2} \right) \approx \frac{10}{3} - \frac{M_0^2}{m_W^2}$. Thus, the lower bound results in:
 \begin{align}
  \frac{\Gamma \left( l_{\alpha} \rightarrow l_{\beta} \gamma \right)}{\Gamma \left( l_{\alpha} \rightarrow l_{\beta} \bar{\nu}_{\beta} \nu_{\alpha} \right)} \geq \frac{3 \alpha}{32 \pi} \frac{M_0^2}{m_W^2} \left( \mathcal{E} \right)_{\alpha \beta}^2
 \end{align}
In this case, the lower bound is additionally suppressed by the small factor $\frac{M_0^2}{m_W^2}$. 
\end{itemize}
Using the experimental values for the branching ratios of the processes $l_{\alpha} \rightarrow l_{\beta} \bar{\nu}_{\beta} \nu_{\beta}$, see e.g. \cite{Beringer:1900zz}, it is found an expression for the branching ratios of the three processes:
\begin{align}
Br_{\mu e} \leq \frac{3 \alpha}{8 \pi} \left( \mathcal{E} \right)_{\mu e}^2, \quad \quad Br_{\tau e} \leq \frac{1}{5.6} \frac{3 \alpha}{8 \pi} \left( \mathcal{E} \right)_{\tau e}^2 \quad \text{and} \quad Br_{\tau \mu} \leq \frac{1}{5.9} \frac{3 \alpha}{8 \pi} \left( \mathcal{E} \right)_{\tau \mu}^2 
\end{align} 
In the limit of a small $\delta \Phi$, $\mathcal{E}_{\alpha \beta}$ is given in leading order in $\delta \Phi$ by:
\begin{align}
\frac{\mathcal{E}_{\alpha \beta}}{c_{\alpha} c_{\beta}} &= \frac{\left( \pi R Y \right)^2}{2 \sin \left( y \right)^2}  \left[ 1 - (q-1) \cos \left( 2 y q - 2 \Phi \right) + q \cos \left( 2 \left[ y - qy + \Phi \right] \right) \right] \\
&= \left( \pi R Y \right)^2 h \left( q,y,\Phi \right), \label{eq:Eab}
\end{align}
where $q=\frac{a}{\pi R}$ and $y=\pi M_0 R$. Remarkably, the only dependence on the flavor is given by the factors $c_{\alpha}$, $c_{\beta}$. Thus, the ratio of two different branching ratios of rare lepton decays is to leading order in $\delta \Phi$ given by $\frac{Br_{\alpha \beta}}{Br_{\gamma \delta}} = \frac{c_{\alpha}^2 c_{\beta}^2}{c_{\gamma}^2 c_{\delta}^2}$. \\
Note that the ratios of the LFV decays in leading order $\delta \phi$ are independent of any simplifications of the loop function, e.g. $F(x_k)=F(M_0)$ for $k \geq 4$. This is the case since, as shown in appendix \ref{App:C}, $U_{\alpha k} U_{k \beta}^{\dagger} = K c_{\alpha} c_{\beta} + \mathcal{O} \left( \delta \phi \right)$ holds. Consequently, the only flavor dependent terms, the factors $c_{\alpha} c_{\beta}$, can be pulled out of the sum in equation (4.2) and therefore the ratios of the decay rates in leading order $\delta \phi$ are independent of the approximations adopted in the loop functions. The results for all four cases for the ratios $\frac{Br_{\tau \mu}}{Br_{\mu e}}=\frac{c_{\tau}^2}{5.9}$ and $\frac{Br_{\tau e}}{Br_{\mu e}} = \frac{c_{\tau}^2}{5.6 c_{\mu}^2}$ are presented in table \ref{tab:ratios}. According to this, there is no reason to expect larger rates for the LFV $\tau$ decays than for the LFV $\mu$ decays. For the IO $Br_{\tau \mu}$ and $Br_{\tau e}$ can even be expected to be one to two orders of magnitude smaller than $Br_{\mu e}$. \\
Furthermore, the same ratios can be expected for lepton flavor violating $Z$ decays, as well, since any one loop diagram contributing to $Z \rightarrow l_{\alpha} l_{\beta}$ includes a factor of $|\sum_{k=1}^{\infty} U_{\alpha k} U_{k \beta}^{\dagger} F_Z\left( x_k \right)|^2$, where $F_Z\left( x_k \right)$ is the loop function of the respective diagram. As for the rare lepton decays, the only dependence on the flavor originates from the mixing matrix elements, which are in leading order in $\delta \Phi$ proportional to $c_{\alpha} c_{\beta}$.     \\
\begin{table}
\centering
\begin{tabular}{| c | c c | c c |} \hline
Case & $\frac{Br_{\tau \mu}}{Br_{\mu e }}$ analytic & $3 \sigma$ & $\frac{Br_{\tau e}}{Br_{\mu e }}$ analytic & $3 \sigma$  \\ \hline
I & $\frac{\cot \left( \Theta_{13} \right)^2 \cos \left( \Theta_{23} \right)^2}{5.9}$ & $[2.36 , 5.53]$ & $\frac{\cot \left( \Theta_{23} \right)^2}{5.6}$ & $[0.10 , 0.29]$ \\ \hline
II & No analytic expression & $[0.10 , 0.42]$ & No analytic expression & $[0.06 , 2.77]$ \\ \hline 
III & $\frac{cos \left( \Theta_{23} \right)^2 \left[ \sin \left( \Theta_{13} \right) \tan \left( \Theta_{13} \right) - \tan \left( \Theta_{23} \right) \right]}{ 5.9 cos \left( \Theta_{13} \right)^2 \tan \left( \Theta_{12} \right)^2}$ &$[0.09 , 0.27]$ & $\frac{1}{5.6}\left( \frac{\tan \left( \Theta_{23} \right) - \sin \left( \Theta_{13} \right) \tan \left( \Theta_{12} \right)}{1+ \sin \left( \Theta_{13} \right) \tan \left( \Theta_{12} \right) \tan \left( \Theta_{23} \right)} \right)^2$ & $[0.07 , 0.23]$ \\ \hline
IV & $\frac{\cos \left( \Theta_{23} \right)^2 \left[ \sin  \left( \Theta_{13} \right) - \tan \left( \Theta_{12} \right) \tan \left( \Theta_{23} \right) \right]^2 }{5.9 \cos \left( \Theta_{13} \right)^2} $ & $[0.01 , 0.04]$ & $ \frac{1}{5.6}\left( \frac{\sin \left( \Theta_{13} \right) - \tan \left( \Theta_{12} \right) \tan \left( \Theta_{23} \right)}{\tan \left( \Theta_{12} \right) + \sin \left( \Theta_{13} \right) \tan \left( \Theta_{23} \right)} \right)^2$ & $[0.03,0.15]$ \\ \hline
\end{tabular}
\caption{Analytic expressions (LO in $\delta \Phi$) for the ratios of the branching ratios of the rare lepton decays in terms of the mixing angles and their $3 \sigma$ regions for the different cases. Case I and II correspond to NO and Case III and IV to IO.   } \label{tab:ratios}
\end{table}
\FloatBarrier
If eq. \eqref{eq:Eab} is combined with eq. \eqref{eq:l-} and $\lambda_3 \approx \sqrt{\Delta m_{\text{atm}}^2} \equiv m_{\nu}$ is assumed, for the branching ratio one obtains:
\begin{align}
 \frac{3 \alpha \left( m_{\nu} \pi R \right)^2 c_{\mu}^2}{32 \pi \, s \left( c_{\mu}, c_{\tau} \right)^2} \left[ \frac{10}{3} - F \left( \frac{M_0^2}{m_W^2} \right) \right]^2 \frac{h \left( q, y, \Phi \right)^2}{f \left( q, y, \Phi \right)^2} \leq Br_{\mu e} \leq \frac{3 \alpha \left( m_{\nu} \pi R \right)^2 c_{\mu}^2}{8 \pi \, s \left( c_{\mu}, c_{\tau} \right)^2} \frac{h \left( q, y, \Phi \right)^2}{f \left( q, y, \Phi \right)^2}. \label{eq:muebounds}
\end{align}
In figure \ref{fig:megwoED}, the upper and the lower bound are presented for two configurations of the parameters $q$, $y$ and $\Phi$ as well as a numerically obtained value for $Br_{\mu e}$. For $M_0 \gtrsim m_W$ the lower bound approaches the upper bound and the numerical value is almost exact. The numerical value for $Br_{\mu e}$ differs significantly from the bounds for $M_0 \lesssim m_W$. It reaches its maximum at roughly $M_0 = m_W$ and decreases slowly afterwards. The maximum value can be estimated by evaluating the lower bound at $M_0=m_W$:
\begin{align}
 Br_{\mu e} \approx \frac{3 \alpha}{128 \pi} \frac{c_{\mu}^2}{s \left( c_{\mu}, c_{\tau} \right)^2} \left( \frac{m_{\nu}}{m_W} \right)^2 y^2 \frac{h \left( q, y, \Phi \right)^2}{f \left( q, y, \Phi \right)^2} \approx 2 \times 10^{-31}  y^2 \frac{h \left( q, y, \Phi \right)^2}{f \left( q, y, \Phi \right)^2} \, .
\end{align}  
In the last step, we adopted the best fit values for scenario I for $c_{\mu}$ and $c_{\tau}$ (see table \ref{tab:par1}). 
If the branching ratio $Br_{\mu e}$ is analyzed for different values of $q$, $y$ and $\Phi$, it is found that $Br_{\mu e}$ lies far below the experimental bounds for most of these values since the ratio $\frac{h \left( q, y, \Phi \right)^2}{f \left( q, y, \Phi \right)^2}$ is not much larger than one. This case is illustrated on the left panel of figure \ref{fig:megwoED}. However, there are configurations of $q$, $y$ and $\Phi$ where the factor $h^2 f^{-2}$ can enhance the branching ratio for this process significantly.
In order to generate a $Br_{\mu e}$ close to the experimental bounds, the factor $h^2 f^{-2}$ is required to be roughly $10^{17}$. \\
The factor is divergent for three different configurations of the parameters $q$, $y$ and $\Phi$. Concerning a large $Br_{\mu e}$, a small deviation from these divergent configurations $\epsilon$ is needed and the dependence on $\epsilon$ is shown in table \ref{tab:epsilon}. Additionally, the influence of these configurations on the value for the phase shift $\delta \Phi$ is presented, which was required to be small. It can be estimated by the ratio of the non-vanishing eigenvalues $\lambda_2$ \eqref{eq:l+} and $\lambda_3$ \eqref{eq:l-}, which has to be $\sim 1$ for the IO and $\sim 5$ or $\sim 0.2$ for the NO, respectively. \\
To conclude, the extra dimensional setup allows for $Br_{\mu e}$ close to experimental limits even with 5D Yukawa couplings of $\mathcal{O} \left( 1 \right)$, if the phases of the Yukawa couplings are close to $\Phi^F \approx \frac{2n+1}{2} \pi + qy$ or $\Phi = \frac{2n+1}{2} \pi + qy -y$. \\
\begin{table}
\centering
\begin{tabular}{|c |c | c |} \hline
Configuration & $h^2 f^{-2}$ & $\delta \Phi$ \\ \hline
$y=0$ & $\epsilon^{-2}$ & $\cos \left( \Phi \right)^2 \epsilon^{-1}$ \\ 
$\Phi = \frac{2n+1}{2} \pi + qy$ & $q^2 \epsilon^{-2}$ & $\epsilon$ \\ 
$\Phi = \frac{2n+1}{2} \pi + qy -y$ & $\left( q-1 \right)^2 \epsilon^{-2}$ & $\epsilon$ \\ \hline 
\end{tabular}
\caption{Dependence of $h^2 f^{-2}$ and $\delta \Phi$ on the deviation $\epsilon$ of the presented configurations. Since $h^2 f^{-2}$ is always $\sim \epsilon^{-2}$, $\epsilon \approx 10^{-7}$ is required to generate $Br_{\mu e} \approx 10^{-13} - 10^{-14}$. Consequently, this would lead to a large $\delta \Phi$ for the first case, which is not desirable since $\delta \Phi \ll 1$ was assumed before. Note that it could be rescued by $\Phi \approx \frac{\pi}{2}$. The remaining two cases have $\delta \Phi \sim \epsilon$, thus leading to a very small $\delta \Phi$ for a large $Br_{\mu e}$. The $\epsilon$ dependence of $\delta \Phi$ is calculated by taking the ratio of $\lambda_2$ and $\lambda_3$, leading to $\delta \Phi \sim f$. } \label{tab:epsilon}
\end{table}
Furthermore, it is possible to extract some information about the fundamental scale of gravity $M_F$. As we figured out in Chapter \ref{sec:mass+pmns}, for 5D Yukawa couplings of $\mathcal{O} \left( 1 \right)$ the neutrino mass scale is given by \eqref{eq:limitseesaw},\eqref{eq:nuscale}: 
\begin{align}
m_{\nu} = \frac{s \left( c_{\mu}, c_{\tau} \right) v^2}{4 M_F} f \left( q,y,\Phi \right) \, . 
\end{align}
In case of a large $Br_{\mu e}$, it is $f \left( q,y,\Phi \right) \approx \epsilon$ and $\frac{h \left( q, y, \Phi \right)^2}{f \left( q, y, \Phi \right)^2} \approx \epsilon^{-2}$, with $\epsilon \ll 1$. Assuming $M_0 > m_W$ and combining the expressions for the neutrino mass scale with the expression for $Br_{\mu e}$ \eqref{eq:muebounds} yields:
\begin{align}
 m_{\nu} = \sqrt{\frac{3 \alpha}{8 \pi}} \frac{c_{\mu}}{s \left( c_{\mu} , c_{\tau} \right)} \frac{m_{\nu} \pi R}{\sqrt{Br_{\mu e}}} \frac{s \left( c_{\mu} , c_{\tau} \right)}{4} \frac{v^2}{M_F} \, .
\end{align}
Rewriting the radius $R$ in terms of the fundamental scale of gravity $M_F$ and the number of extra dimensions $n$, that are experienced by gravity, and using eq. \eqref{eq:RMF} one finds a lower bound on $M_F$ in terms of $n$:
\begin{align}
 M_F = \left( \sqrt{\frac{3 \alpha c_{\mu}^2}{512 \pi  Br_{\mu e}}} v^2 M_P^{\frac{2}{n}} \right)^{\frac{n}{2 \left( n + 1 \right)}} \geq \left( 1.33 \times 10^9 \, \mathrm{GeV}^2 M_P^{\frac{2}{n}} \right)^{\frac{n}{2 \left( n + 1 \right)}}
\end{align} 
For some values of $n$, the lower limit of $M_F$ is presented in table \ref{tab:llimits}. Note that for $n \rightarrow \infty$ the limit approaches $M_F \geq 36.5 \, \mathrm{TeV}$. \\
Likewise, one finds a lower limit on the inverse radius $R^{-1}$ in terms of the number of extra dimensions, which results in $R^{-1} \geq 5.6 \, \mathrm{GeV}$ for $n=2$. Since the KK neutrino mass is to a good approximation given by $M_k = y \left( \pi R \right)^{-1} + k R^{¬1} $ with $|y| \leq \frac{\pi}{2}$ the heavier mass eigenstates corresponding to the KK neutrinos, in most cases, cannot be produced in kaon or muon decays. Note that at least the lightest one could be produced for $y \ll 1$.  
\begin{table}
\centering
\begin{tabular}{|c |c c c c c|} \hline
$n$ & $2$ & $6$ & $10$ & $20$ & $50$  \\ \hline
Lower Limit on $M_F$  & $1.5 \, \mathrm{EeV}$ & $3.4 \, \mathrm{PeV}$ & $660 \, \mathrm{TeV}$ & $166 \, \mathrm{TeV}$ & $68 \, \mathrm{TeV}$  \\ 
Lower Limit on $R^{-1}$  & $5.6 \, \mathrm{GeV}$ & $2.4 \, \mathrm{TeV}$ & $12.7 \, \mathrm{TeV}$ & $50.3 \, \mathrm{TeV}$ & $123 \, \mathrm{TeV}$  \\ \hline 
\end{tabular}
\caption{Lower Limits for the fundamental scale of gravity $M_F$ for different number of extra dimensions $n$. } \label{tab:llimits}
\end{table}

\FloatBarrier
\section{Summary and Conclusions}
In this paper, we have studied an extra dimensional seesaw mechanism with a single right handed bulk neutrino. The SM particles are confined to a 4D brane. Shifting the brane away from the orbifold fixed points allows to generate two non-vanishing mass-squared differences as required by neutrino oscillation experiments. \\
In particular, we have worked out the flavor structure without adopting a non-unitarity approximation of the $3 \times 3$ submatrix. This allows us to study the phenomenological consequences of the right handed bulk neutrino. \\
In a first step, we studied the neutrino mass generation and mixing. We further simplified the analysis by assuming CP conversation and that the ratios of the Yukawa coupling of the $Z_2$ even component and $Z_2$ odd component of the right handed neutrino to the SM neutrinos $\frac{h_2^l}{h_1^l}$ are almost the same for all three generations. The allowed parameter space is presented in table \ref{tab:par1}. Additionally, the model predicts one massless neutrino which can be probed in large scale structure surveys in cosmology.    \\
It is pointed out that the model is capable of generating $Br_{\mu e}$ close to the experimental bounds even with Yukawa couplings close to one. As discussed in section \ref{sec:UVandPheno}, the contribution to $l_i \rightarrow l_j \gamma$ is maximized if the lightest KK excitation has roughly the W-Boson mass. Due to the suppression of the Yukawa coupling by the extra dimension it is still possible to generate the observed neutrino mass with a Yukawa coupling of order one in this case. However, this effect is not strong enough to produce $Br_{\mu e}$ close to $10^{-13}$. Therefore, some fine tuning of the brane shift, the ratio of the lowest KK mass to $R^{-1}$ and $\frac{h_2^l}{h_1^l}$ is necessary. Note that this behavior is not an exclusive feature of the brane shifted model and is also possible without a brane shift. In this case, $M_0$ close to $\frac{1}{2} R^{-1}$ is required to generate a sizable $Br_{\mu e}$. However, the brane shift is necessary to generate two neutrino mass squared differences. \\
A strong prediction of the model within the approximations mentioned above are the ratios of flavor violating charged lepton decay and Z decay branching ratios which are correlated with the neutrino mixing angles and the neutrino mass hierarchy. Thus, the model could be tested by the next generation of experiments looking for charged lepton flavor violation. Furthermore, it could allow for a distinction of the neutrino mass hierarchies by the measurement of lepton flavor violating processes. \\
Finally, note that the model might also be probed in neutrino oscillation experiments due to effects of non-unitarity \cite{FernandezMartinez:2007ms}, although these effects are not further investigated within this work.       


\appendix

\section{Solution of the infinite sums} \label{sec:appA}
In equation \eqref{eq:CP} sums as e.g.
\begin{align}
A^{F_1} A^{F_2}\sum \limits_{k=-\infty}^{\infty} \frac{\cos \left( \frac{ka}{R} + \Phi^{F_1} \right) \cos \left( \frac{ka}{R} + \Phi^{F_2} \right)}{M_0 + \frac{k}{R} - \lambda} \equiv S \left( F_1 , F_2 \right) = S \left( F_2 , F_1 \right). \label{eq:sum}
\end{align}
have to be solved. To solve the sum a method is used similar to that in \cite{Bhattacharyya:2002vf}. The key point is to write the brane shift $a$ in a way that $\frac{a}{\pi R}$ becomes a rational number.
\begin{align}
a = \frac{r \pi R}{q} \quad r,q \in \mathbb{N} \quad \text{and} \quad q>r
\end{align}
For the following calculation $r=1$ is chosen, but the calculation works in a similar way with $r \neq 1$. The periodicity of the Yukawa couplings to the KK modes is used to split the infinite sum over $k$ into two sums, one infinite sum of $n$ and one finite sum of $l$. \\
The relation between the old and new summation variables is $k=qn+l$. Since a step in $n$ causes a step of $q$ in $k$, the second sum over $l$ has to be introduced. This sum has to fill the gaps between a given $k$ and $k+q$. Hence, this sum has to run from $l=0$ to $l=q-1$. Thus results in:
\begin{align*}
\frac{S \left( F_1, F_2 \right)}{A^{F_1} A^{F_2}} &=  \sum \limits_{l=0}^{q-1} \sum \limits_{n=-\infty}^{\infty} \frac{\cos \left( n \pi + \frac{l}{q} \pi + \Phi^{F_1} \right) \cos \left( n \pi + \frac{l}{q} \pi + \Phi^{F_2} \right)}{M_0+\frac{qn}{R}+\frac{l}{R} - \lambda} \\
\Rightarrow \frac{S \left( F_1, F_2 \right)}{A^{F_1} A^{F_2}} &= \sum \limits_{l=0}^{q-1} \sum \limits_{n=-\infty}^{\infty} \frac{\cos \left(  \frac{l}{q} \pi + \Phi^{F_1} \right) \cos \left( \frac{l}{q} \pi + \Phi^{F_2} \right)}{M_0+\frac{qn}{R}+\frac{l}{R} - \lambda} \\
\Rightarrow \frac{S \left( F_1, F_2 \right)}{A^{F_1} A^{F_2}} &= \sum \limits_{l=0}^{q-1} \cos \left(  \frac{l}{q} \pi + \Phi^{F_1} \right) \cos \left( \frac{l}{q} \pi + \Phi^{F_2} \right) \sum \limits_{n=-\infty}^{\infty} \frac{1}{M_0+\frac{qn}{R}+\frac{l}{R} - \lambda}.
\end{align*}
In the calculation above $\frac{ka}{R} = \frac{qn+l}{R}a = n \pi + \frac{l}{q} \pi$ is used. In the next step, profit is made of the periodicity of the cosine function. By using the periodicity the dependence of the numerator of $n$ is eliminated. Consequently, the numerator can be pulled out of the sum over $n$. \\
With this it is possible to solve the infinite sum of $n$:
\begin{align*}
 \sum \limits_{n=-\infty}^{\infty} \frac{1}{B+\frac{qn}{R}} = \frac{1}{B} + \sum \limits_{n=1}^{\infty} \left( \frac{1}{B+\frac{qn}{R}} + \frac{1}{B-\frac{qn}{R}} \right) = \frac{1}{B} + \sum \limits_{n=1}^{\infty} \frac{2 B}{B^2 - \frac{q^2}{R^2}n^2}
\end{align*}
where $B=M_0+\frac{l}{R}-\lambda$ holds. \\
Comparing the result with the series representation of $\cot \left( x \right)$ leads to the following result: 
\begin{align*}
\frac{1}{B} + \sum \limits_{n=1}^{\infty} \frac{2 B}{B^2 - \frac{q^2}{R^2}n^2} = \frac{R}{q}\pi \cot \left( \frac{R}{q}\pi B \right).
\end{align*}
Thus it is possible to write the sum as:
\begin{align*}
\frac{S \left( F_1, F_2 \right)}{A^{F_1} A^{F_2}} &= \sum \limits_{l=0}^{q-1} \cos \left(  \frac{l}{q} \pi + \Phi^{F_1} \right) \cos \left( \frac{l}{q} \pi + \Phi^{F_2} \right) \frac{R}{q}\pi \cot \left( \frac{\pi R \left[ M_0 - \lambda \right]}{q} + \frac{l}{q} \pi  \right)\\ 
&= \frac{R}{q} \pi \sum \limits_{l=0}^{q-1} \cos \left(  \frac{l}{q} \pi + \Phi^{F_1} \right) \cos \left( \frac{l}{q} \pi + \Phi^{F_2} \right) \frac{\cos \left( \frac{\Theta}{q} + \frac{l}{q} \pi \right)}{\sin \left( \frac{\Theta}{q} + \frac{l}{q} \pi \right)},
\end{align*}
where $\Theta=\pi R \left( M_0 - \lambda \right)$ holds. 
The finite sum of $l$ remains:
\begin{align*}
 \frac{R \pi}{q } \sum \limits_{l=0}^{q-1} \left[ \cos \left( 2 \frac{l}{q} \pi + \Phi^{F_1} + \Phi^{F_2} \right) + \cos \left( \Phi^{F_1} -\Phi^{F_2} \right) \right] \cos \left( \frac{\Theta}{q} + \frac{l}{q} \pi \right)  \frac{\prod \limits_{ m \neq l}^{q-1} \sin \left( \frac{\Theta}{q} + \frac{m}{q} \pi \right)}{\prod \limits_{k=0}^{q-1} \sin \left( \frac{\Theta}{q} + \frac{k}{q} \pi \right)}.  
\end{align*}
In this form it is possible to exploit the following relations, which are similarly used in \cite{Bhattacharyya:2002vf}. It is made reference to the fact that the proof is long and mainly relies on some properties of the unit roots $z^q=1$ like $\sum \limits_{\text{all roots}}  z = 0$ and that their total product is $\left( -1 \right)^{q-1}$ : 
\begin{align*}
&\prod \limits_{k=0}^{q-1} \sin \left( \frac{\Theta}{q} + \frac{k}{q} \pi \right) = 2^{1-q} \sin \Theta \\
&\sum \limits_{l=0}^{q-1} \cos \left( \frac{\Theta}{q} + \frac{l}{q} \pi \right) \prod \limits_{m \neq l}^{q-1} \sin \left( \frac{\Theta}{q} + \frac{m}{q} \pi \right) = 2^{1-q} q \cos \Theta  \\
&\sum \limits_{l=0}^{q-1} \cos \left( 2 \frac{l}{q} \pi + \Phi^{F_1} + \Phi^{F_2} \right) \cos \left( \frac{\Theta}{q} + \frac{l}{q} \pi \right) \prod \limits_{ m \neq l}^{q-1} \sin \left( \frac{\Theta}{q} + \frac{m}{q} \pi \right) = \\ &2^{1-q} q \cos \left( \Phi^{F_1} + \Phi^{F_2} + \frac{q-2}{q} \Theta \right).
\end{align*} 
This relations lead to:
\begin{align*}
 \frac{S \left( F_1, F_2 \right)}{A^{F_1} A^{F_2}} = \frac{\pi R}{2} \left[ \frac{\cos \left( \Phi^{F_1} + \Phi^{F_2} + \frac{q-2}{q} \Theta \right)}{\sin \Theta} + \cos \left( \Phi^{F_1} - \Phi^{F_2} \right) \cot \Theta \right].
\end{align*}
$q = \frac{\pi R}{a}$ is resubstituted what leads to the final result:
\begin{align}
  \frac{S \left( F_1 , F_2 \right)}{\pi R A^{F_1} A^{F_2}} &=  \left[ \cot \left(  \pi R \left[M_0 - \lambda \right] \right) \cos \left( \Phi^{F_1} - a \left[M_0 - \lambda \right] \right) \cos \left( \Phi^{F_2} - a \left[M_0 - \lambda \right] \right) \right. \nonumber \\ &- \left.\frac{1}{2} \sin \left( \Phi^{F_1} + \Phi^{F_2} - 2a \left[M_0 - \lambda \right] \right) \right]. \label{eq:sum1}
\end{align}  
The second sum, which is to solve, is:
\begin{align}
A^{F_1} A^{F_2}\sum \limits_{k=-\infty}^{\infty} \frac{\cos \left( \frac{ka}{R} + \Phi^{F_1} \right) \cos \left( \frac{ka}{R} + \Phi^{F_2} \right)}{\left( M_0 + \frac{k}{R} - \lambda \right)^2} = S_2 \left( F_1 , F_2 \right). \label{eq:sum2}
\end{align}
The solution is obtained by differentiating $S\left( F_1, F_2 \right)$ with respect to $\Theta$. 
\begin{align*}
&\frac{d}{d \Theta} S\left( F_1, F_2 \right) = \frac{d}{d \Theta} A^{F_1} A^{F_2}\sum \limits_{k=-\infty}^{\infty} \frac{\cos \left( \frac{ka}{R} + \Phi^{F_1} \right) \cos \left( \frac{ka}{R} + \Phi^{F_2} \right)}{\frac{k}{R} + \frac{\Theta}{\pi R}} \\
&= -\frac{A^{F_1} A^{F_2}}{\pi R}\sum \limits_{k=-\infty}^{\infty} \frac{\cos \left( \frac{ka}{R} + \Phi^{F_1} \right) \cos \left( \frac{ka}{R} + \Phi^{F_2} \right)}{\left( \frac{k}{R} + \frac{\Theta}{\pi R} \right)^2} = -\frac{1}{\pi R} S_2 \left( F_1 , F_2 \right) \\
\Rightarrow & S_2 \left( F_1 , F_2 \right) = - \pi R \frac{d}{d \Theta} S\left( F_1, F_2 \right).
\end{align*}
The derivative with respect to $\Theta$ is performed. This leads to the following result:
\begin{align}
&\frac{S_2 \left( F_1 , F_2 \right)}{\pi^2*R^2 A^{F_1} A^{F_2}} =  \frac{ \cos \left( \Phi^{F_1} - a \left[M_0 - \lambda \right] \right) \cos \left( \Phi^{F_2} - a \left[M_0 - \lambda \right] \right)}{\sin \left( \pi R \left[ M_0- \lambda \right] \right)^2} - \nonumber \\
&\frac{a}{\pi R} \cot \left( \pi R \left[ M_0- \lambda \right] \right) \left( \frac{\cos \left( \Phi^{F_1} - a \left[M_0 - \lambda \right] \right)}{\sin \left( \Phi^{F_2} - a \left[M_0 - \lambda \right] \right)} +  \frac{\cos \left( \Phi^{F_2} - a \left[M_0 - \lambda \right] \right)}{\sin \left( \Phi^{F_1} - a \left[M_0 - \lambda \right] \right)} \right) \nonumber  \\
&\frac{a}{\pi R} \cos \left( \Phi^{F_1} + \Phi^{F_2} - 2 a \left[ M_0 - \lambda \right] \right). \label{eq:sum2sol}
\end{align} 
\section{PMNS Matrix} \label{sec:appB}
In this section, the relations for the mixing matrix and its unitarity violation are quickly derived. It is assumed that $A R \ll 1$ holds, what leads to $Y \mathcal{M}^{-1} \ll 1$. Additionally, all Yukawa couplings are considered to be real. In this limit, it can be assumed that the mass matrix is diagonalized by:
\begin{align}
 \mathcal{U} = \begin{pmatrix}
                \mathcal{U}_P & A \\
                B & 1 
               \end{pmatrix}
\end{align}
Since the overall mixing matrix $U$ should be unitary, i.e. $\mathcal{U} \mathcal{U}^T = \mathcal{U}^T \mathcal{U} = 1$, $B^T = -\mathcal{U}_P^T A$ and $\mathcal{U}_P \mathcal{U}_P^T = 1 - A A^T$ hold. It is obtained:
\begin{align}
 \mathcal{U}^T \begin{pmatrix}
               0 & Y^T \\
               Y & \mathcal{M}
               \end{pmatrix} \mathcal{U} = \begin{pmatrix}
               \mathcal{U}_P^T Y^T B + B^T Y \mathcal{U}_P + B^T \mathcal{M} B & \mathcal{U}_P^T Y^T + B^T Y A + B^T \mathcal{M} \\
               Y \mathcal{U}_P + A^T Y^T B + \mathcal{M} B & A^T Y^T + Y A + \mathcal{M}  
               \end{pmatrix}. \label{eq:B1}
\end{align}
Since $U$ should diagonalize the mass matrix the off diagonal components have to vanish. Substituting $B^T = -\mathcal{U}_P^T A$ into the off diagonal components yields the following condition:
\begin{align}
A \left( M + Y A \right) = Y^T . \label{eq:B2}
\end{align}
For the case $Y \mathcal{M}^{-1} \ll 1$, the mixing is expected to be small and therefore, the lower right component of \eqref{eq:B1} is approximately $\mathcal{M}$, leading to a small $YA$ compared to $\mathcal{M}$.
Employing $YA \ll \mathcal{M}$ in equation \eqref{eq:B2}, results in:
\begin{align}
 A = Y^T \mathcal{M}^{-1}.
\end{align}
Consequently, the upper left component of the matrix in equation \eqref{eq:B1} simplifies to:
\begin{align}
\mathcal{U}_P^T \left( - Y^T \mathcal{M}^{-1} Y \right) \mathcal{U}_P. 
\end{align}
Therefore, if $\mathcal{U}_P$ diagonalizes the matrix $- Y^T \mathcal{M}^{-1} Y$, its deviation of unitarity is given by:
\begin{align}
 \mathcal{U}_P \mathcal{U}_P^T = 1 - Y^T \mathcal{M}^{-2} Y.
\end{align}
The combination $\mathcal{U}_P \mathcal{U}_P^T$ is of greater interest than $\mathcal{U}_P^T \mathcal{U}_P$ since the influence on lepton flavor violation is our main interest and for that an expression for $\sum_i \left( U_P \right)_{Fi} \left( U_P \right)_{iF^{'}} = \mathcal{U}_P \mathcal{U}_P^T$ is needed.

\section{Flavor Ratios} \label{App:C}
In this section, we will show that the ratios of two different LFV decays, e.g. $\l_{\alpha} \rightarrow l_{\beta} \gamma$, is given by the ratio of the corresponding product $c_{\alpha} c_{\beta}$. Neglecting phase space effects, the flavor dependence originates from the following factor:
\begin{align}
\frac{|\sum_{k=1}^{\infty} U_{\alpha k} U_{k \beta}^{\dagger} F\left( x_k \right)|^2}{\left( U U^{\dagger} \right)_{\alpha \alpha} \left( U U^{\dagger} \right)_{\beta \beta} }. 
\end{align}
Here, $F(x)$ is some loop function and $x$ is a function of the masses of the particles propagating in the loop. In section \ref{sec:UVandPheno}, a lower bound was derived by assuming that all KK particles have the same mass. However, this approximation is not necessary in order to obtain the flavor ratios in leading order $\delta \phi$. For that, it is inevitable to calculate the mixing matrix elements $U_{\alpha k}$ explicitly. Therefor, we have to find the eigenvectors of \eqref{eq:MMatrix}. For the components of the $k$-th eigenvector it is obtained:
\begin{align}
 v_k^i = - \frac{\sum_F m_i^F v_k^F}{M_0 + i R^{-1} - \lambda_k} \quad \text{and} \quad v_k^F = \frac{1}{\lambda_k} \sum \limits_{i=-\infty}^{\infty} m_i^F v_k^i \, .
\end{align}
The lower index represents the $k$-th mass eigenstate and $|\lambda_k|<|\lambda_{k+1}|$ holds. Consequently, $v_{1,2,3}$ correspond to the active neutrino mass eigenstates. The upper index represents the flavor eigenstates with $F=e,\mu,\tau$ and $i \in [-\infty,\infty]$. Therewith, the mixing matrix elements are given by:
\begin{align}
 U_{\alpha k} = \frac{v_k^\alpha}{\sqrt{ \sum \limits_F \left( v_k^F \right)^2 + \sum \limits_i \left( v_k^i \right)^2 }} \, . 
\end{align}
Combining the equations above allows to write the product $U_{\alpha k} U_{\beta k}^{\dagger}$ as:
\begin{align}
U_{\alpha k} U_{\beta k}^{\dagger} &= \frac{v_k^{\alpha} v_k^{\beta}}{\sum \limits_F \left( v_k^F \right)^2 + \sum \limits_i \left( v_k^i \right)^2 } \\
&= \frac{1}{\lambda_k^2 } \frac{1}{\sum \limits_F \left( v_k^F \right)^2 + \sum \limits_i \left( v_k^i \right)^2} \sum \limits_{F_1,F_2} S \left( \alpha , F_1, \lambda_k \right) S \left( \beta, F_2, \lambda_k \right) v_k^{F_1} v_k^{F_2} \, .
\end{align}
Note that the sums $S \left( \alpha, \beta \right)$ \eqref{eq:sum1} are in leading order $\delta \phi$ given by:
\begin{align}
S \left( \alpha , \beta, \lambda_k \right) = c_{\alpha} c_{\beta} K_k + \mathcal{O} \left( \delta \phi \right) \, ,
\end{align}
where $K_k$ is a function of parameters of the model which do not depend on the Flavor. Consequently, the only flavor dependence is given by:
\begin{align}
U_{\alpha k} U_{\beta k}^{\dagger} = \frac{1}{\lambda_k^2 } \frac{K_k^2 c_{\alpha} c_{\beta}}{\sum \limits_F \left( v_k^F \right)^2 + \sum \limits_i \left( v_k^i \right)^2} \sum \limits_{F_1,F_2} c_{F_1} c_{F_2} v_k^{F_1} v_k^{F_2} \, = K' c_{\alpha} c_{\beta} .
\end{align}
Moreover, we can approximate $\left( U U^{\dagger} \right)_{\alpha \alpha}$ with 1 since the deviation from unitarity is expected to be small and $x (1-x)^{-2} \approx x$ holds for $x \ll 1$.
\bibliography{references}

\begin{thebibliography}{10}

\bibitem{ArkaniHamed:1998rs}
N.~Arkani-Hamed, S.~Dimopoulos, and G.~R. Dvali,
\newblock Phys. Lett. {\bf B429}, 263 (1998), hep-ph/9803315.

\bibitem{Randall:1999vf}
L.~Randall and R.~Sundrum,
\newblock Phys. Rev. Lett. {\bf 83}, 4690 (1999), hep-th/9906064.

\bibitem{ArkaniHamed:1998vp}
N.~Arkani-Hamed, S.~Dimopoulos, G.~R. Dvali, and J.~March-Russell,
\newblock Phys. Rev. {\bf D65}, 024032 (2001), hep-ph/9811448.

\bibitem{Dienes:1998sb}
K.~R. Dienes, E.~Dudas, and T.~Gherghetta,
\newblock Nucl.Phys. {\bf B557}, 25 (1999), hep-ph/9811428.

\bibitem{Bhattacharyya:2002vf}
G.~Bhattacharyya, H.~V. Klapdor-Kleingrothaus, H.~Pas, and A.~Pilaftsis,
\newblock Phys. Rev. {\bf D67}, 113001 (2003), hep-ph/0212169.

\bibitem{Grossman:1999ra}
Y.~Grossman and M.~Neubert,
\newblock Phys.Lett. {\bf B474}, 361 (2000), hep-ph/9912408.

\bibitem{Pilaftsis:1999jk}
A.~Pilaftsis,
\newblock Phys. Rev. {\bf D60}, 105023 (1999), hep-ph/9906265.

\bibitem{Ioannisian:1999cw}
A.~Ioannisian and A.~Pilaftsis,
\newblock Phys. Rev. {\bf D62}, 066001 (2000), hep-ph/9907522.

\bibitem{Lukas:2000rg}
A.~Lukas, P.~Ramond, A.~Romanino, and G.~G. Ross,
\newblock JHEP {\bf 04}, 010 (2001), hep-ph/0011295.

\bibitem{Antusch:2003kp}
S.~Antusch, J.~Kersten, M.~Lindner, and M.~Ratz,
\newblock Nucl. Phys. {\bf B674}, 401 (2003), hep-ph/0305273.

\bibitem{Antusch:2014woa}
S.~Antusch and O.~Fischer,
\newblock JHEP {\bf 10}, 094 (2014), 1407.6607.

\bibitem{Accomando:1999sj}
E.~Accomando, I.~Antoniadis, and K.~Benakli,
\newblock Nucl. Phys. {\bf B579}, 3 (2000), hep-ph/9912287.

\bibitem{Donini:1999px}
A.~Donini and S.~Rigolin,
\newblock Nucl. Phys. {\bf B550}, 59 (1999), hep-ph/9901443.

\bibitem{DREINER20101}
H.~K. Dreiner, H.~E. Haber, and S.~P. Martin,
\newblock Physics Reports {\bf 494}, 1  (2010).

\bibitem{Antoniadis:1998ig}
I.~Antoniadis, N.~Arkani-Hamed, S.~Dimopoulos, and G.~R. Dvali,
\newblock Phys. Lett. {\bf B436}, 257 (1998), hep-ph/9804398.

\bibitem{Aad:2015zva}
ATLAS, G.~Aad {\em et~al.},
\newblock Eur. Phys. J. {\bf C75}, 299 (2015), 1502.01518,
\newblock [Erratum: Eur. Phys. J.C75,no.9,408(2015)].

\bibitem{Gonzalez-Garcia:2015qrr}
M.~C. Gonzalez-Garcia, M.~Maltoni, and T.~Schwetz,
\newblock (2015), 1512.06856.

\bibitem{Berryman:2016szd}
J.~M. Berryman, A.~de~Gouvea, K.~J. Kelly, O.~L.~G. Peres, and Z.~Tabrizi,
\newblock Phys. Rev. {\bf D94}, 033006 (2016), 1603.00018.

\bibitem{Antusch:2006vwa}
S.~Antusch, C.~Biggio, E.~Fernandez-Martinez, M.~B. Gavela, and J.~Lopez-Pavon,
\newblock JHEP {\bf 10}, 084 (2006), hep-ph/0607020.

\bibitem{Beringer:1900zz}
Particle Data Group, J.~Beringer {\em et~al.},
\newblock Phys. Rev. {\bf D86}, 010001 (2012).

\bibitem{FernandezMartinez:2007ms}
E.~Fernandez-Martinez, M.~B. Gavela, J.~Lopez-Pavon, and O.~Yasuda,
\newblock Phys. Lett. {\bf B649}, 427 (2007), hep-ph/0703098.

\end{thebibliography}
\bibliographystyle{h-physrev}

\end{document}